\documentclass[acmtog]{acmart}
\usepackage{booktabs} % For formal tables
\usepackage{caption,subcaption}
\usepackage{hyperref}
\AtBeginShipout{%
	\ifnum\value{page}>1 %
	\typeout{* Additional boxing of page `\thepage'}%
	\setbox\AtBeginShipoutBox=\hbox{\copy\AtBeginShipoutBox}%
	\fi
}

\usepackage[ruled]{algorithm2e} % For algorithms

\SetAlFnt{\small}
\SetAlCapFnt{\small}
\SetAlCapNameFnt{\small}
\SetAlCapHSkip{0pt}
\hypersetup{draft}

\begin{document}

% Title portion
\title{Dynamic Upsampling of Smoke through Dictionary-based Learning} 

\author{Kai Bai}
\affiliation{
	\institution{ShanghaiTech University}}
\email{baikai@shanghaitech.edu.cn}

\author{Wei Li}
\affiliation{
	\institution{ShanghaiTech University}}
\email{liwei@shanghaitech.edu.cn}

\author{Mathieu Desbrun}
\affiliation{
	\institution{Caltech/ShanghaiTech University}}
\email{mathieu@caltech.edu}

\author{Xiaopei Liu}
\affiliation{
	\institution{ShanghaiTech University}}
\email{liuxp@shanghaitech.edu.cn}

\renewcommand\shortauthors{Bai, K. et al}

\begin{teaserfigure}
	\centering\includegraphics[width=0.98\textwidth]{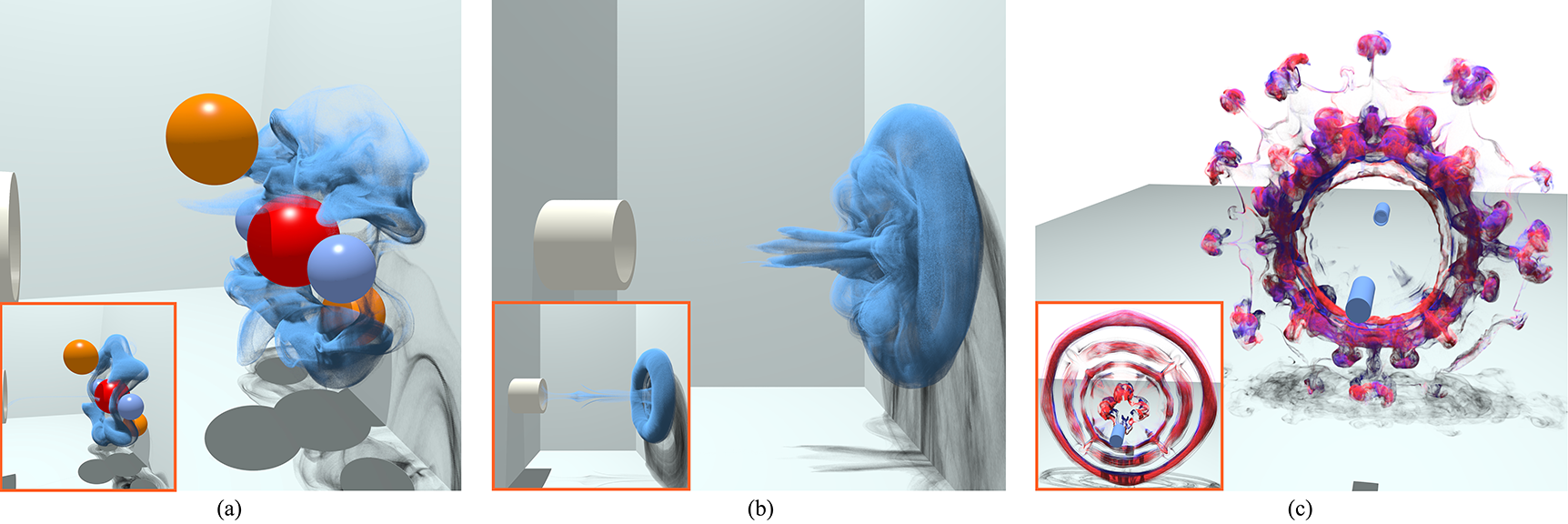}\vspace*{-4mm}
	\caption{\textbf{Fine smoke animations from low-resolution simulation inputs.} We present a versatile method for dynamic upsampling of smoke flows which handles a variety of animation contexts: (a) \emph{generalized synthesis}, where a few pairs of low-res and high-res simulations are used as a training set (here, a smoke simulation with a single ball obstacle, and another one with 5 balls arranged in a vertical plane), from which one can synthesize a high-res simulation from any coarse input (here, 5 balls at a $45^{\circ}$ angle compared to the second training simulation); (b) \emph{restricted synthesis}, where the training set contains only a few sequences (here, 4 simulations with various inlet sizes), leading to faster training and more predictive synthesis results (the input simulation uses an inlet size not present in the training set); and (c) \emph{re-simulation}, where the training set consists of only one simulation, from which we can quickly ``re-simulate'' the original flow very realistically for small variations of the original animation (here, vortex rings colliding). Note that all our synthesized high-resolution smoke animations share strong similarities with their corresponding fine numerical simulations, even if they are generated over an order of magnitude faster. \vspace*{5mm}}
	\label{fig:teaser}
\end{teaserfigure}

\begin{abstract}
	\label{sec:abstract}
	% 200 words limit
	%LONG VERSION
	\hspace{-1.3mm}
	Simulating turbulent smoke flows is computationally intensive due to their intrinsic multiscale behavior, thus requiring relatively high resolution grids to fully capture their complexity. 
	For iterative editing or simply faster generation of smoke flows, dynamic upsampling of an input low-resolution numerical simulation is an attractive, yet currently unattainable goal. 
	In this paper, we propose a novel dictionary-based learning approach to the dynamic upsampling of smoke flows. 	
	For each frame of an input coarse animation, we seek a sparse representation of small, local velocity patches of the flow based on an over-complete dictionary, and use the resulting sparse coefficients to generate a high-resolution smoke animation sequence. 
	We propose a novel dictionary-based neural network which learns both a fast evaluation of sparse patch encoding and a dictionary of corresponding coarse and fine patches from a sequence of example simulations computed with any numerical solver. 
	Our upsampling network then injects into coarse input sequences physics-driven fine details, unlike most previous approaches that only employed fast procedural models to add high frequency to the input. 
	We present a variety of upsampling results for smoke flows and offer comparisons to their corresponding high-resolution simulations to demonstrate the effectiveness of our approach.
	\vspace*{-2mm}
	
\end{abstract}


\keywords{Fluid Simulation, Dictionary Learning, Neural Networks, Smoke Animation\vspace*{-2mm}}

\maketitle

\section{Introduction}
\label{sec:intro}

Visual simulation of smoke is notoriously difficult due to its highly turbulent nature, resulting in vortices spanning a vast range of space and time scales.
As a consequence, simulating the dynamic behavior of smoke realistically requires not only sophisticated non-dissipative numerical solvers~\cite{Kim-2005,Selle-2008,Zhang-2015}, but also a spatial discretization with sufficiently high resolution to capture fine-scale structures, either uniformly~\cite{Kim-2008b,Zehnder-2018} or adaptively~\cite{Losasso-2004,Weissmann:2010,Zhang-2016}.
This inevitably makes such direct numerical simulations computationally intensive.\\[-3mm]

%\paragraph{Fine simulations from coarse inputs.}
In order to compromise between efficiency and visual realism for large scale scenes, the general concept of physics-inspired upsampling of dynamics~\cite{Kavan-11-CUP} can be leveraged:
low-resolution simulations can be computed first, from which a highly-detailed flow is synthesized using fast procedural models that are only loosely related with the underlying fluid dynamics, e.g., noise-based~\cite{Kim-2008a,Bridson-2007} or simplified turbulence models~\cite{Schechter-2008,Pfaff-2010}.
Very recently, machine learning has even been proposed as a means to upsample a coarse flow simulation~\cite{Chu-2017} (or even a downsampled flow simulation~\cite{Xie-2018,Werhahn-2019}) to obtain finer and more visually-pleasing results inferred from a training set of actual simulations. However, while current upsampling methods can add visual complexity to a coarse input, the synthesized high-resolution fluid flow often fails to exhibit the correct fine details that the original physical equations are expected to give rise to: 
the inability to properly capture small-scale vortical structures leads to visual artifacts, making the resulting flows not quite realistic.\\[-3mm]

\begin{figure}[t]
	\centering
	\includegraphics[width=0.95\columnwidth]{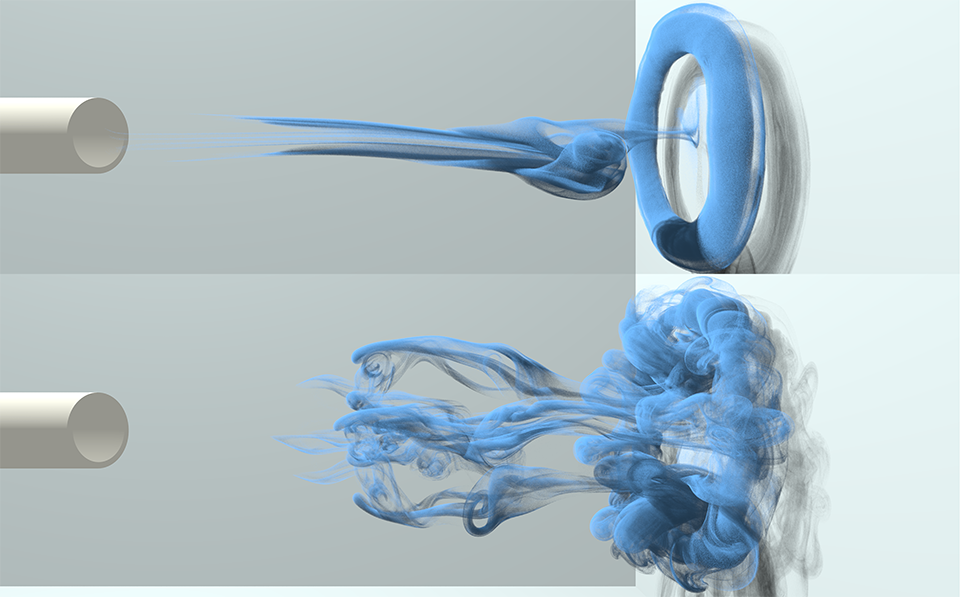}\vspace{-3mm}
	\caption{\textbf{Coarse vs. fine smoke simulations.} A smoke simulation computed using a low (top: $50\times75\times50$) vs. a high resolution (bottom: $200\times300\times200$) respectively, for the same Reynolds number ($5000$). Flow structures are visually quite distinct since different resolutions resolve different physical scales, thus producing quite different instabilities.	\vspace{-5mm}}
	\label{fig:high_low_example}
\end{figure}

%\paragraph{Dictionary learning of simulation upsampling.}
Synthesizing a high-resolution flow from a low-resolution one is fundamentally difficult because their global and local structures can differ significantly (see Fig.~\ref{fig:high_low_example} for an example).
Discrepancy between low-res and high-res numerical simulations is not only due to discretization errors, but also to flow instabilities, becoming more pronounced for high Reynolds number flows. However, one could hope that many of the details of the flow can be inferred by comparing local patches of the input coarse flow to a catalog of existing low-res/high-res pairs of numerical simulations: a proper local encoding of existing upsampled sequences could help predict the appearance and evolution of fine structures from existing coarse input simulations --- in essence, learning how the natural shedding of small vortices appears from the corresponding coarse simulation. 
In this paper, we propose to upsample smoke motions through \emph{dictionary learning}~\cite{Garcia-2018} (a common approach in image upsampling~\cite{Yang-2010}) based on the observation that although turbulent flows look complex, local structures and their evolution do not differ significantly as they adhere to the self-advection process prescribed by the fluid equations: local learning through sparse coding followed by synthesis through a dictionary-based neural network is thus more physically appropriate than global learning methods such as convolutional neural networks~\cite{Tompson-2016}.

\subsection{Related Work}
\label{sec:related_work}
Smoke animation has been widely studied for more than two decades in computer graphics.
We review previous work relevant to our contributions, covering both traditional simulation of smoke and data-driven approaches to smoke animation.\\[-6mm]

\paragraph{Numerical smoke simulation.}
Smoke animation has relied most frequently on numerical simulation of fluids in the past. Fast fluid solvers~\cite{Stam-1999}, and their higher-order~\cite{Kim-2005,Selle-2008}, momentum-preserving~\cite{Lentine-2010} or advection-reflection~\cite{Zehnder-2018} variants, can efficiently simulate smoke flows on uniform grids. However, creating complex smoke animations requires relatively high resolutions to capture fine details. Unstructured grids~\cite{Klingner-2006,De-2015,Mullen-2009,Ando-2013} and adaptive methods, where higher resolutions are used in regions of interest and/or with more fluctuations~\cite{Losasso-2004,Zhu-2013,Setaluri-2014} have been proposed to offer increased efficiency --- but the presence of smoke turbulence in the entire domain often prevents computational savings in practice.
On the other hand, particle methods, e.g, smoothed particle hydrodynamics~\cite{Desbrun:1996:SP,Becker-2007,Solenthaler-2009,Akinci-2012,Ihmsen-2014,Peer-2015,Winchenbach-2017} and power particles~\cite{De-2015} can easily handle adaptive simulations.
However, a large number of particles are necessary to obtain realistic smoke animations to avoid poor numerical accuracy for turbulent flows.
Hybrid methods~\cite{Zhu-2005,Raveendran-2011,Jiang-2015,ZhangXX-2014,Zhang-2016}, which combine both particles and grids, can be substantially faster but particle-grid interpolations usually produce strong dissipation unless polynomial basis functions are used for improved numerics~\cite{Fu-2017}. These methods remain very costly in the context of turbulent smoke simulation.
Another set of approaches that can simulate smoke flow details efficiently are vortex methods~\cite{Park-2005,Golas-2012}; in particular, vortex filaments~\cite{WeiBmann-2010} and vortex sheets~\cite{Brochu-2012,Pfaff-2012} are both effective ways to simulate turbulent flows for small numbers of degrees of freedom, and good scalability can be achieved with fast summation techniques~\cite{ZhangXX-2014}. However, no existing approach has been proposed to upsample low-resolution vortex-based simulations to full-blown high-resolution flows to further accelerate fluid motion generation. We note finally that a series of other numerical methods have been developed to offer efficiency through the use of other fluid models~\cite{Chern-2016} or of massively-parallelizable mesoscopic models like the lattice Boltzmann method~\cite{Chen-1998,DHumieres-2002,Geier-2006,Liu-2012,Daniel-2014,Rosis-2017,Li-2018}, but here again, the ability to run only a coarse simulation to quickly generate a high-resolution fluid motion has not been investigated. \\[-4mm]

\paragraph{Early upsampling attempts.} Over the years, various authors have explored ways to remediate the shortcomings induced by numerical simulation on overly coarse grids in the hope of recovering high-resolution results. Reinjecting fine details through vorticity confinement~\cite{John-1994,Fedkiw-2001}, Kolmogorov-driven noise~\cite{Bridson-2007,Kim-2008b}, vorticity correction~\cite{Zhang-2015}, smoke style transfer~\cite{Sato-2018} or modified turbulence models~\cite{Schechter-2008,Pfaff-2010} partially helps, but visually important vortical structures are often lost: none of these approaches provides a reliable way to increase the resolution substantially without clearly deviating from the corresponding simulation on a fine computational grid. \\[-4mm]

\paragraph{Data-driven approaches}
Given the computational complexity of smoke animation through numerical simulation, data-driven approaches have started to emerge in recent years as a promising alternative. 
Some techniques proposed generating a flow field entirely based on a trained network, completely avoiding numerical simulation for fluid flows~\cite{Jeong-2015,Guo-2016,Wiewel-2018,Kim-2018}. Other data-driven methods either try to solve fluid flow equations more efficiently, (e.g., the work of~\cite{Tompson-2016,Umetani-2018} trains a neural network to quickly predict pressure without solving the Poisson equation),
or directly synthesize flow details for smoke and liquid animations from low-resolution simulations, (e,g.,~\cite{Chu-2017,Werhahn-2019,Xie-2018} create high frequency smoke details based on neural networks, while \cite{Um-2018} models fine-detail splashes for liquid simulations from existing data).
Yet, these recent data-driven upsampling approaches do not generate turbulent smoke flows that are faithful to their physical simulations using similar boundary conditions: the upsampling of a coarse motion fails to reconstruct physically-meaningful (and more importantly, visually-expected) details, e.g., vortex ring dynamics, leapfrogging phenomena, etc. Our work focuses on addressing this deficiency via a novel neural network based on dictionary learning. \\[-4mm]

\subsection{Overview}
Although smoke flows exhibit intricate structures as a whole, the short-term evolution of a \textit{local patch} of smoke is in fact quite simple, following a restricted gamut of behaviors. 
The complexity of high resolution turbulent flow fields mainly comes from \textit{the rich combination of these local motions}, only constrained to respect a global incompressibility condition. Given that coarse simulations already enforce incompressibility, this patch-based view of the motion motivates the idea of \emph{dictionary learning}, as used in image upsampling, to achieve physically-driven upsampling for coarse smoke flows.\\[-4mm]

However, existing dictionary learning methods for image upsampling cannot be directly used for flow synthesis. First and foremost, we have to learn structures from vector fields (or vortex fields) instead of scalar fields, a much richer set than typical image upsampling methods are designed for. Second, we are dealing with a dynamical system instead of a static image, so we must also adapt the actual upsampling process appropriately. \\[-2mm]

In this paper, we propose a novel network structure for dictionary learning of smoke flow upsampling, where the integration of fine Navier-Stokes flow patches is learned from their coarse versions. We ensure good spatial and temporal coherency by directly learning from the high-resolution residuals between coarse motion predictions and actual fine motion counterparts.  
Plausible high-resolution flows can then be quickly synthesized from low-resolution simulations, providing much better approximation to the real fine-scale dynamics than existing data-driven methods, and with computing efficiency often over an order of magnitude faster than direct simulation depending on the resolution for upsampling; for instance, Fig.~\ref{fig:teaser} shows examples of animation results generated from three types of upsampling based on our dictionary learning approach, where coarse simulations are upsampled by a factor of 64 (4$\times$4$\times$4) using different training sets, all revealing the typical vortex structures that a (significantly more) costly fine simulation would exhibit. We demonstrate that our approach provides a visually-plausible interpolation between (and to a certain extent, extrapolation from) training examples to produce upsampling of a coarse smoke flow simulation.
We also evaluate our results in terms of visual quality, energy spectrum, synthesis error compared to fine simulations, as well as computing performance in order to thoroughly validate the advantages of our method.\\[-4mm]
\section{Background on Dictionary Learning}
\label{sec:overview}

We first review traditional dictionary learning for image up-sampling, as its foundations and algorithms will be key to our contributions once properly adapted to our animation context.

\begin{figure}[t]
	\centering
	\includegraphics[width=.96\columnwidth]{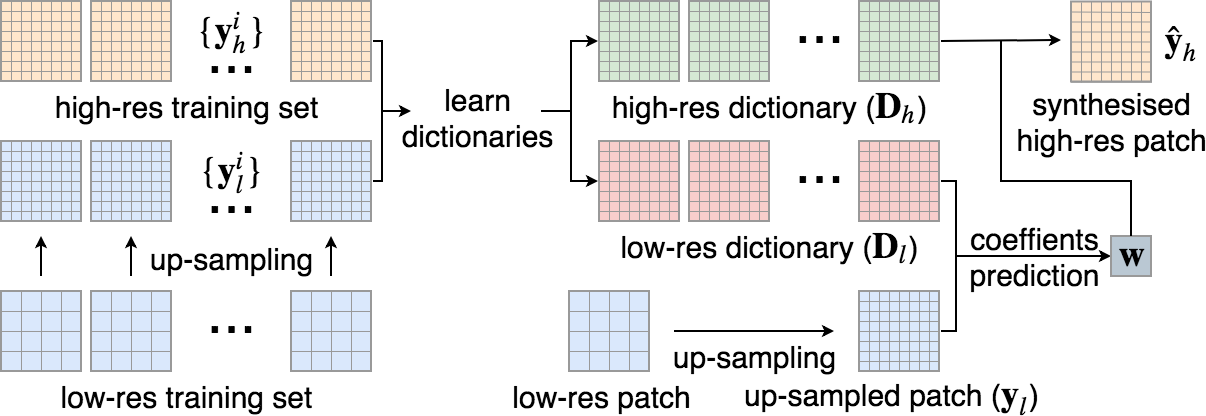}\vspace*{-2mm}
	\caption{\textbf{Dictionary learning for image upsampling.} In order to synthesize high-resolution images, one prepares a training set of local patch pairs ($\mathbf{y}_l^i$ and $\mathbf{y}_h^i$) from low and high resolution images respectively (left), from which we can learn two dictionaries ($\mathbf{D}_l$ and $\mathbf{D}_h$). Given a low resolution image, each coarse patch is then used to predict a set of sparse coefficients $\mathbf{w}$ such that the corresponding patch in the high-resolution image can be synthesized using $\mathbf{D}_h$ and the same sparse coefficients $\mathbf{w}$.\vspace*{-4mm}}
	\label{fig:traditional_learning_diagram}
\end{figure}

\subsection{Foundations}

In image upsampling, a high-resolution image is synthesized from a low-resolution image with a learned dictionary of local patches as summarized in Fig.~\ref{fig:traditional_learning_diagram}: the input low-resolution image is first written as a sparse weighted sum of local ``coarse'' patches; then the resulting high-resolution image is written as the weighted sum, with exactly the same weights, of the corresponding ``fine'' (upsampled) patches, when each corresponding pair of coarse and fine patches comes from a training set of upsampled examples. 
One thus needs to find a dictionary of patch pairs, and a way to write any low-resolution image as a linear combination of coarse dictionary patches.

\paragraph{Role of a dictionary.} 
A dictionary for image upsampling is a set of upsampled low-resolution local patches $\{\smash{\mathbf{d}_i^l}\}_i$ of all the same size, and their associated high-resolution local patches $\{\smash{\mathbf{d}_i^h}\}_i$ (for instance, all of the size 5$\times$5 pixels).
By storing the dictionary patches as vectors, any upsampled coarse patch $\smash{\mathbf{y}_l}$ can be approximated by a patch $\smash{\tilde{\mathbf{y}}_l}$ that is a sparse linear combination of coarse dictionary patches, i.e., $\smash{\tilde{\mathbf{y}}_l\!=\!\sum_i w_i\mathbf{d}_l^i}$ with a sparse set of coefficients $w_i$.
An upsampled patch $\smash{\tilde{\mathbf{y}}_h}$ corresponding to the input upsampled coarse patch $\smash{\mathbf{y}_l}$ can then be expressed as $\smash{\tilde{\mathbf{y}}_h\!=\!\sum_i w_i\mathbf{d}_h^i}$. 
For convenience, we will denote by $\smash{\mathbf{D}_l}$ the matrix storing all the coarse dictionary patches $\smash{\mathbf{D}_l=\bigl(\mathbf{d}_1^h ... \mathbf{d}_N^h\bigr)}$ (where each patch is stored as a vector), and similarly for all the high-resolution dictionary patches using $\smash{\mathbf{D}_h}$ --- such that a patch $\smash{\tilde{\mathbf{y}}_l}$ (resp., $\smash{\tilde{\mathbf{y}}_h}$) can be concisely computed as $\smash{\mathbf{D}_l} \mathbf{w}$ (resp., $\smash{\mathbf{D}_h} \mathbf{w}$) where $\mathbf{w}\!=\!(w_1, ..., w_N)$.

\paragraph{Finding a dictionary.} For a given training set of coarse and fine image pairs, we can find a dictionary with $N$ patch elements by making sure that it not only captures all coarse patches well, but its high-resolution synthesis also best matches the input fine images. 
If we denote by $\mathbf{Y}_l$ the vector storing all the coarse patches $\smash{\mathbf{y}_l}$ available from the training set and by $\mathbf{Y}_h$ the vector of their corresponding fine patches $\mathbf{y}_h$, the dictionaries as well as the sparse weights are found through a minimization~\cite{Yang-2010}:\vspace*{-0.5mm}
\begin{equation} \underset{\mathbf{D}_l, \mathbf{D}_h, \mathbf{w}}{argmin}\;\;\frac{1}{2} \left\| \mathbf{Y}_l-\widetilde{\mathbf{Y}}_l(\mathbf{D}_l) \right \|_{2}^{2} + \left\| \mathbf{Y}_h-\widetilde{\mathbf{Y}}_h(\mathbf{D}_h) \right \|_{2}^{2} + \lambda \left \| \mathbf{w} \right \|_{1} \; , \label{eq:traditional_dl_objective} \vspace*{-0.5mm}
\end{equation}
which evaluates the representative power of the coarse dictionary patches (through the first term) and the resulting upsampling quality (using the $\ell_2$ difference between the upsampled patches $\smash{\widetilde{\mathbf{Y}}_h}$ and their ground-truth counterparts $\smash{\mathbf{Y}_h}$) while penalizing non-sparse weights via an $\ell_1$ norm of $\mathbf{w}$ times a regularization factor $\lambda$. 
Solving for the dictionary patches minimizing this functional is achieved using the K-SVD method~\cite{Aharon-2006}. 

\paragraph{Upsampling process.}
Once the training stage has learned a dictionary, then upsampling a low-resolution input image is done by finding a sparse linear combination of the coarse dictionary patches for each local patch of the input. The method of orthogonal matching pursuit (OMP) is typically used to find the appropriate (sparse) weights that best reproduce a local patch based on the dictionaries (other pursuit methods used in graphics can potentially be used too~\cite{Von-2013,Teng-2015}), from which the high-resolution patch is directly reconstructed using these weights, now applied to the high-resolution dictionary patches.
The final high-resolution simulation frame is generated by blending all locally synthesized high-resolution patches together.

%Then, given any low resolution field, we can optimize for $\mathbf{w}$ at each local patch using the method of orthogonal matching pursuit (OMP)~\cite{Tropp-2004} with already known $\mathbf{D}_l$, such that the high resolution patch can be obtained by using the same $\mathbf{w}$ but with high resolution dictionary $\mathbf{D}_h$.

\subsection{Scalable solver to find sparse linear combinations}
\begin{figure}[t]
	\centering
	\includegraphics[width=1.0\columnwidth]{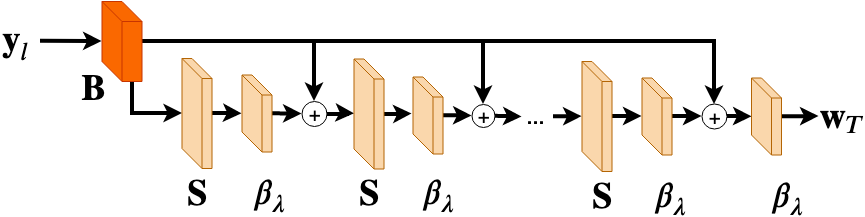}\vspace*{-3mm}
	\caption{\textbf{Original LISTA network.} Illustration of feed-forward LISTA neural network~\protect\cite{Gregor-2010} with $T$ layers.\vspace*{-4mm}}
	\label{fig:LISTA_network}
\end{figure}

\begin{figure*}[t]
	\centering
	\includegraphics[width=1.0\textwidth]{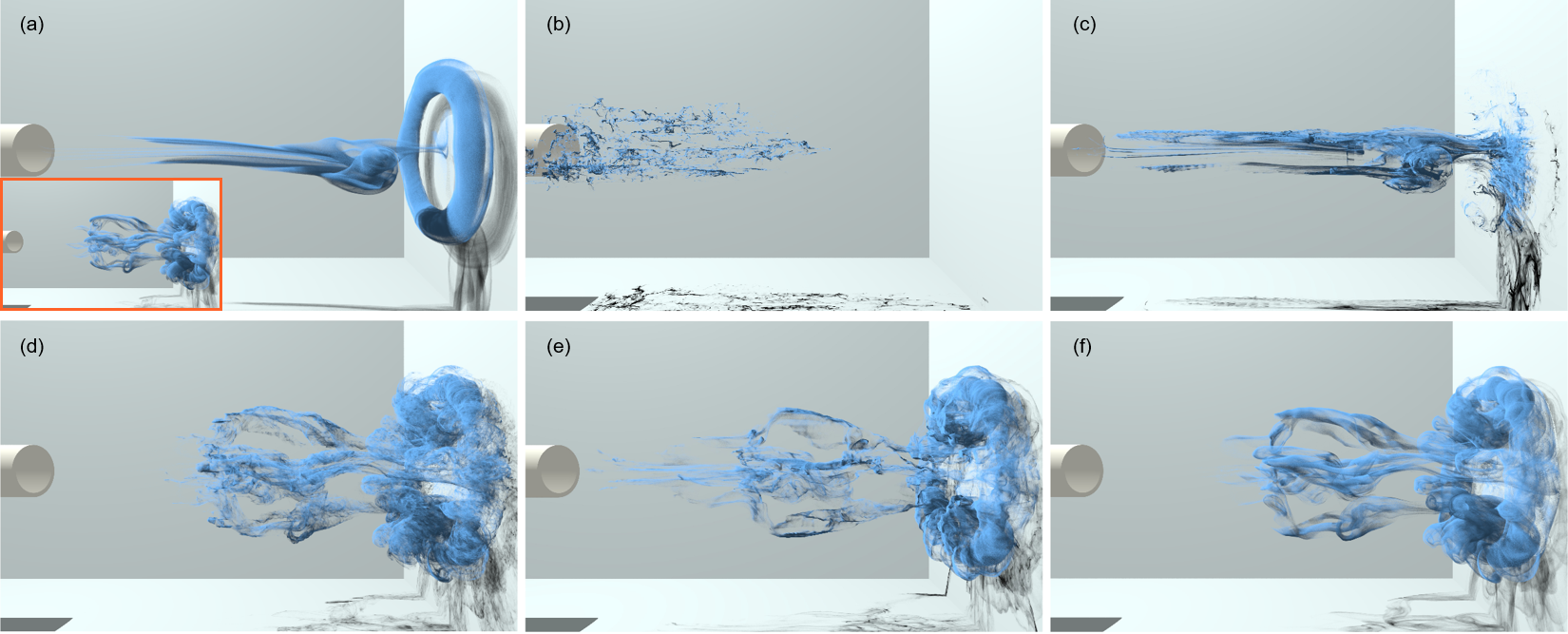}\vspace*{-3mm}
	\caption{\textbf{Strategies for upsampling.} From an input low-resolution simulation in (a) (its fine simulation counterpart is shown as inset), the most straightforward method to upsample it is to use the traditional dictionary learning from image up-sampling, this time in 3D, using either the OMP method (b) or a LISTA network (c) --- but because of how different low and high resolution simulations look, they both produce poor synthesis results; reformulating the problem as a dictionary learning reconstruction, we can obtain much better result in (d), although it tends to be slightly noisy; with our novel network and by representing each local patch with velocities only (e), the results exhibit spatial and temporal incoherence; when adding to each local patch spatial and time codes (f), a coherent synthesis can be obtained, looking quite close to the fine simulation counterpart.\vspace*{-2mm}}
	\label{fig:overview_compare}
\end{figure*}

Looking ahead at our task at hand, the fact that we will have to deal with 3D vector-based fields will make the dimension of the vectors $\mathbf{y}_l$ quite larger than what is typically used in image upsampling. In this case, the OMP-based optimization required to perform upsampling will become extremely slow, and may even return poor results as shown in Fig.~\ref{fig:overview_compare}(b).
Thus, a more efficient method to compute $\mathbf{w}$ given a low-resolution patch is required.
One scalable approach is the LISTA neural network proposed by Gregor and LeCun~\cite{Gregor-2010}: it is a learning-based formulation derived from the iterative shrinkage thresholding algorithm (ISTA)~\cite{Daubechies-2004} using the iteration process:\vspace*{-0.5mm}
\begin{equation}
\mathbf{w}_{t+1} = \beta \bigl( \mathbf{S}\mathbf{w}_{t}+\mathbf{B}\mathbf{Y}\,;\lambda \bigr) \; , \label{eq:w_iteration}\vspace*{-0.5mm}
\end{equation}
where $\mathbf{B}\!=\!h\mathbf{D}^T$, $\mathbf{S}\!=\!\mathbb{I}-\mathbf{B}\mathbf{D}$ ($\mathbb{I}$ being the identity matrix) with $\mathbf{D}\!=\! [\mathbf{D}_l; \mathbf{D}_h]^T$ the matrix concatenating the coarse and fine dictionary patches and $\mathbf{Y}\!=\! [\mathbf{Y}_l; \mathbf{Y}_h]^T$; $h$ is the iteration step size, and $\beta(\cdot\,;\cdot)$ is a vector function constructed to enforce the sparsity of $\mathbf{w}$:\vspace*{-0.5mm}
\begin{equation}
\beta_i(\mathbf{x}\,;\lambda) = \operatorname{sgn}(x_{i}) \;\operatorname{max}\left\{ \left |x_{i}\right |-\lambda,0 \right \} \; , \label{eq:beta_theta}\vspace*{-0.5mm}
\end{equation}
where $\beta_i$ is the $i$-th component of the output vector from the vector function $\beta$; $x_{i}$ represents the $i$-th component of vector $\mathbf{x}$.
This immediately corresponds to a feed-forward neural network with $T$ layers, see Fig.~\ref{fig:LISTA_network}, where $t$ in Eq.~\eqref{eq:w_iteration} denotes the $t$-th layer in the network, and $\beta$ is the activation function of the network.

In the original LISTA algorithm, in order to enable better prediction of the final $\mathbf{w}$, $\mathbf{B}$ and $\mathbf{S}$ are both updated during learning while $\lambda$ remains fixed.
The number of layers in the network corresponds to the total number of iterations in Eq.~\eqref{eq:w_iteration}, with more layers typically leading to more accurate prediction. The network is trained using a set of pairs $\{(\mathbf{y}_i,\mathbf{w}_i), i=1,2,...,k.\}$, whose weights are computed using the OMP method on a given learned dictionary $\mathbf{D}$, and with a loss function for the network defined as:\vspace*{-1mm}
\begin{equation}
L_{T}(\Theta )=\frac{1}{k}\sum_{i=1}^{k}\left \| \mathbf{w}_{T}(\mathbf{y}_{i};\Theta )-\mathbf{w}_{i} \right \|_{2}^{2} \; , \label{eq:loss_lista}\vspace*{-1mm}
\end{equation}
where $T$ is the index of the output layer, and $\Theta$ is the vector assembled from all parameters (including $\mathbf{B}$, $\mathbf{S}$
and $\lambda$ from Eq.~\eqref{eq:w_iteration}).
After training, we can use the learned $\Theta$ to predict $\mathbf{w}$ from an input $\mathbf{y}_l$ by going through the whole network. This LISTA-based solve provides much higher efficiency than the traditional OMP approach.

\subsection{Inadequacy of direct upsampling on coarse simulation}
\label{sec:inadequate}
While the idea of upsampling a coarse smoke motion in a frame-by-frame fashion through a direct application of the image upsampling approach presented above sounds attractive, it is bound to fail for a number of reasons. First, the issue of coherency over time is not addressed at all: the weights used for a given spatial region over two successive frames have no reasons to bear any resemblance to each other, thus potentially creating motion flickering. Second, one major difference between the image upsampling problem and our smoke upsampling target is that the coarse and fine simulations can differ widely due to the chaotic changes that turbulence naturally exhibits (Fig.~\ref{fig:high_low_example}). Indeed, image upsampling relies on an energy (Eq.~\eqref{eq:traditional_dl_objective}) which puts coarse approximation and fine approximation errors on equal footing, and the LISTA approach based on the iterative optimization of Eq.~\eqref{eq:w_iteration} may also fail to converge to the right OMP solution if $\mathbf{y}_l$ and $\mathbf{y}_h$ differ significantly in structure, since it performs essentially a local search (see Fig.~\ref{fig:overview_compare}(c) for such an experiment). This suggests a change of both the objective for smoke upsampling and the LISTA network formulation. We present our approach next based on the idea that the coarse simulation can be used as \emph{a predictor of the local motion}, from which \emph{the correction needed to get a high resolution frame} is found through dictionary learning. 

\section{Smoke Upsampling via Dictionary Learning}

%From the above analysis, it is clear that LISTA-like network is possible to have good synthesis result using Eq.~\ref{eq:our_dl_objective}, but we need to further relax the constraints it imposes.
%The LISTA network usually requires a pre-optimized dictionary $\mathbf{D}_h$ in Eq.~\ref{eq:our_dl_objective} such that the prediction of $\mathbf{w}(\mathbf{y}_l)$ can then be trained, where the parameters $\mathbf{B}$ and $\mathbf{S}$ are all shared among different layers, although they can be allowed to change over the learning process.
%Since $\mathbf{D}_h$ is obtained from another optimization process which is fixed during network learning, it does not ensure that optimal $\mathbf{w}$ can be found which well minimize Eq.~\ref{eq:our_dl_objective}.
%Although the dictionary can also be involved as parameters during learning, the result does not have obvious improvement.
%In addition, the sharing of parameter $\mathbf{S}$ over different network layers also largely restricts the function spaces the network can produce, which significantly reduces the possibility of the learning process to find a good predictor for $\mathbf{w}$.
%This brings us the idea to construct our new network with much relaxed constraints, which involves $\mathbf{D}_h$ as a parameter to learn, and uses different $\mathbf{S}$ for different layers.

We now delve into our approach for smoke upsampling. 
We provide the intuition behind our general approach, before detailing the new neural network. 
Finally, we discuss how we augment the patch representation to offer a spatially and temporally coherent upsampling of smoke flows, and provide a simple method to have a better training of our upsampling network.

\subsection{Our approach at a glance}
Based on the relevance of dictionary learning to the upsampling of smoke flows, but given the inadequacy of the current techniques to the task at hand, we propose a series of changes in the formulation of the upsampling problem. \\[-5mm]

\paragraph{Prediction/correction.}
In our dynamical context, it is often beneficial to consider a motion as a succession of changes in time. We thus formulate the upsampling of a coarse patch $\mathbf{y}_l$ as a patch $\tilde{\mathbf{y}}_h\!=\!\text{up}(\mathbf{y}_l) +\Delta_h$, where up$(\cdot)$ corresponds to a straighforward spatial upsampling through direct enlargment of the coarse patch, and $\Delta_h$ is a residual high-resolution patch. This amounts to a predictor-corrector upsampling, where the coarse patch is first upsampled straightforwardly by $up(\cdot)$ before details are added. The residual patches should not be expected to only have small magnitudes, though: as we discussed earlier, differences between coarse and fine simulations can be large in turbulent regions of the flow. Since we will use a dictionary of these high-frequency details, their magnitude has no negative influence, only their diversity matters. \\[-10mm]

\paragraph{Residual dictionary.}
Since a smoke flow locally follows the Navier-Stokes equations, we can expect that the residuals can be well expressed by a fine dictionary $\mathbf{D}_h$.
This is indeed confirmed numerically: if one uses K-SVD to solve for a high-resolution dictionary (of 400 patches) with around 3M training patches from a single fine simulation, the dictionary-based reconstruction is almost visually perfect (albeit a little noisier) as demonstrated in Fig.~\ref{fig:overview_compare}(d), confirming that the local diversity of motion is, in fact, limited. We thus expect a residual $\Delta_h$ in our approach to be well approximated by a sparse linear combination of elements of a (high-resolution) dictionary $\mathbf{D}_h$, i.e., a residual is nearly of the form $\Delta_h \!\approx\! \mathbf{D}_h \mathbf{w}$. Just like in the case of image upsampling, sparsity of the weights is preferable as it avoids the unnecessary blurring introduced by the linear combination of too many patches.\\[-5mm]

\paragraph{Variational formulation}
For efficiency reasons, we discussed in Sec.~\ref{sec:inadequate} that using a LISTA-based evaluation of the sparse weights is highly preferable to the use of OMP. This means that we need to train a network to learn to compute, based on a coarse input $\mathbf{y}_l$, the sparse weights $\mathbf{w}(\mathbf{y}_l)$. Thus, in essence, we wish to modify the traditional upsampling minimization of Eq.\eqref{eq:traditional_dl_objective} to instead minimize the errors in reconstruction of the type $\left\| \mathbf{y}_h-\operatorname{up}(\mathbf{y}_l)-\mathbf{D}_h \mathbf{w}(\mathbf{y}_l) \right \|_{2}^{2}$ on a large series of training patches (with control over the sparsity of $\mathbf{w}$) \emph{while also training a LISTA-like network for the weights}. Other notions of reconstruction errors, based on the vorticity, the difference of gradients, or even the divergence of the upsampled patches would also be good to incorporate in order to offer more user control over the upsampling process.\\[-4mm]

Based on these assumptions, we introduce a new neural network design, where learning a (high-resolution residual) dictionary and training a network to efficiently compute sparse linear coefficients are done simultaneously---thus requiring a single optimization. 

\subsection{Neural network design}
\begin{figure}[t]
	\centering
	\includegraphics[width=1.0\columnwidth]{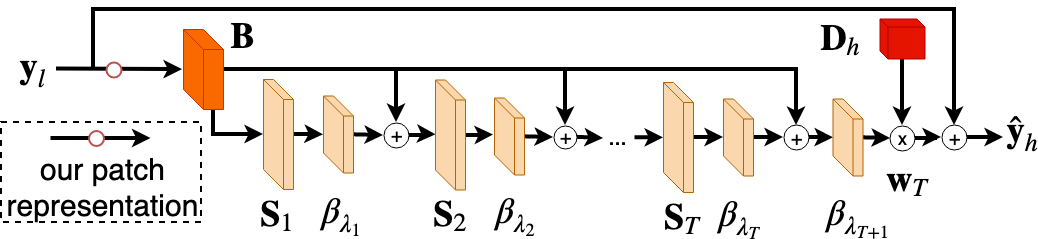}\vspace*{-2mm}
	\caption{\textbf{Our dictionary-based neural network.} We modify the original LISTA network of Fig.~\ref{fig:LISTA_network} by adding layer-specific matrices $\mathbf{S}_t$, regularization parameters $\lambda_t$ as well as the residual dictionary as parameters to learn.\vspace*{-3mm}}
	\label{fig:our_network}
\end{figure}

Our proposed network follows mostly the structure of the LISTA network for sparse coding~\cite{Gregor-2010}, in the sense that it is also composed of several layers representing an iterative approximation of the sparse weights. Two key differences are introduced: first, we add more flexibility to the network by letting each of the $T$ layers not only have its own regularization parameter $\lambda_t$, but also its own matrix $\mathbf{S}_t$; second, while the original LISTA network refers to the the sparse weights $\mathbf{w}$ computed by the OMP method to define the loss, our loss will be measured based on the quality of the reconstruction engendered by the final weights and the dictionary $\mathbf{D}_h$ (the loss will be explicitly provided next in Sec.~\ref{sec:lossDesign}). 
Our fast approximation of sparse coding is achieved through the following modified LISTA-like iteration \vspace*{-1.5mm}
\begin{equation}
\mathbf{w}_{t+1} = \beta \bigl( \mathbf{S}_t\mathbf{w}_{t}+\mathbf{B}\mathbf{y}\,;\lambda_t \bigr) \; , \label{eq:w_iteration_ours}\vspace*{-1.5mm}
\end{equation}
where $\beta$ is the same activation function as in Eq.~\eqref{eq:beta_theta} to enforce sparsity of $\mathbf{w}$. Our novel neural network, summarized in Fig.~\ref{fig:our_network}, can optimize all its network parameters (i.e., the residual dictionary $\mathbf{D}_h$, mapping matrices $\mathbf{S}_t$, regularization parameters $\lambda_t$, and the matrix $\mathbf{B}$) by the standard back-propagation procedure through the $T$ different layers during learning as we will go over later.

\subsection{Loss function design}
\label{sec:lossDesign}
To successfully guide our network during training, an important factor is the choice of loss function. Unlike the 
LISTA network for which the loss function from Eq.~\eqref{eq:loss_lista} requires a set of sparse coefficients $\mathbf{w}$, we construct our loss function directly based on the quality of synthesis results of our network.

\paragraph{$\ell_2$ synthesis error}
One measure for our loss function is the difference between an upsampled patch $\smash{\tilde{\mathbf{y}}^i_h}$ found from a low-resolution patch $\smash{\mathbf{y}^i_l}$ and the ground-truth high-resolution patch $\smash{\mathbf{y}^i_h}$ from a training set containing $K$ patches:\vspace*{-1.5mm}
\begin{equation}
\mathcal{E}_{\ell} = \sum_{i=1}^{K}\left\| \mathbf{y}_h^i-(\mathbf{y}_l^i+\mathbf{D}_h \mathbf{w}_T(\mathbf{y}_l^i;\Theta)) \right \|_{2}^{2} \; , \label{eq:epsilon_l}\vspace*{-1.5mm}
\end{equation}
where $\mathbf{w}_T$ contains the final approximation of the weights since $T$ is the last layer of our network, and the vector $\Theta$ stores all our network parameters ($\mathbf{D}_h$, $\mathbf{B}$, $\mathbf{S}_t$ and $\lambda_t$ for $t\!=\!1...T$).\\[-4mm]

\paragraph{Sobolev synthesis error}
However, using the $\ell_2$ norm measure alone in the loss function is not sufficient to correctly differentiate high frequency structures.
Thus, we also employ the $\ell_2$ norm of the gradient error between synthesized patches and ground-truth patches:\vspace*{-1.5mm}
\begin{equation}
\mathcal{E}_g = \sum_{i=1}^{K}\left\| \nabla[\mathbf{y}_h^i]-\nabla[\mathbf{y}_l^i+\mathbf{D}_h\mathbf{w}_T(\mathbf{y}_l^i;\Theta)] \right \|_{2}^{2} \; . \label{eq:epsilon_g}\vspace*{-1.5mm}
\end{equation}
where $\nabla[\cdot]$ is a component-wise gradient operator defined as $\nabla[\mathbf{x}]=[\nabla x_1, \nabla x_2, ..., \nabla x_n]^T$.\\[-4mm]

\paragraph{Divergence synthesis error}
Since we are synthesizing incompressible fluid flows, it also makes sense to include in the loss function a measure of the divergence error between synthesized and ground-truth patches. While ground-truth patches are divergence-free if a patch representation purely based on the local vector field is used, we will argue that other fields (e.g., vorticity) can be used as well; hence for generality, we use:\vspace*{-2mm}
\begin{equation}
\mathcal{E}_d = \sum_{i=1}^{k}\left\| \nabla\cdot(\mathbf{y}_h^i)-\nabla\cdot(\mathbf{y}_l^i+\mathbf{D}_h\mathbf{w}_T(\mathbf{y}_l^i;\Theta)) \right \|_{2}^{2} \; . \label{eq:epsilon_d}\vspace*{-2mm}
\end{equation}\\[-6mm]

\paragraph{Final form of our loss function}
We express our loss function as: \vspace*{-1.0mm}
\begin{equation}
L_{T}(\Theta)=\alpha_{l}\mathcal{E}_l + \alpha_{g}\mathcal{E}_g + \alpha_{d}\mathcal{E}_d + \alpha_{\Theta}\,\|\Theta\|_{2}^{2} \; ,\vspace*{-1.0mm}
\end{equation}
where the last $\ell_2$ norm on the network parameters helps avoiding over-fitting during learning.
The parameters $\alpha_{l}$, $\alpha_{g}$, $\alpha_{d}$ and $\alpha_{\Theta}$ help balance between training and test losses, and we set them to $\alpha_{l} \!=\! 1$, $\alpha_{g} \!=\! 0.05$, $\alpha_{d} \!=\! 0.05$ and $\alpha_{\Theta} \!=\! 0.5$ in all our training experiments.
One may notice that these values differ significantly, especially for the terms involving gradients; this is because although input velocities are normalized, the gradient values may have much larger ranges, which should be given smaller parameter values.\\[-4mm]

\subsection{Augmented patch encoding}
\label{sec:method}

Until now, we have not discussed what is exactly encoded in a local patch. Since we are trying to upsample a vector field in order to visualize the fine behavior of a smoke sequence, an obvious encoding of the local coarse flow is to use a small patch of coarse velocities, storing the velocities present in an $n\!\times\!n\!\times\!n$ local neighborhood in a vector $\mathbf{y}_l$ of length $N$ where $N\!=\!3n^3$ in this case; a high-resolution patch is similarly encoded, involving a finer subgrid representing the same spatial size as the coarse patch. 
Using our network with such an encoding already performs reasonably well as Fig.~\ref{fig:overview_compare}(e) demonstrates, but the results are not fully spatially and temporally coherent, at times creating visual artifacts. 
Fortunately, we designed our approach to be general so that a number of improvements can be made to remedy this situation. 
Of course, growing the size $n$ of the local patch itself would be one solution, but it would come at the cost of a dramatic increase in computational complexity and learning time, defeating the very purpose of our effort. 
We can, instead, keep the same spatial patch size $n$, but augment the patch with extra data to help further improve spatio-temporal coherence by making the prediction of our residual dictionary less myopic: increasing $N$ helps provide \emph{more dynamic context} for both learning and synthesis of high-resolution smoke flows.\\[-4mm]

\begin{figure}[t]
	\centering
	\includegraphics[width=0.8\columnwidth]{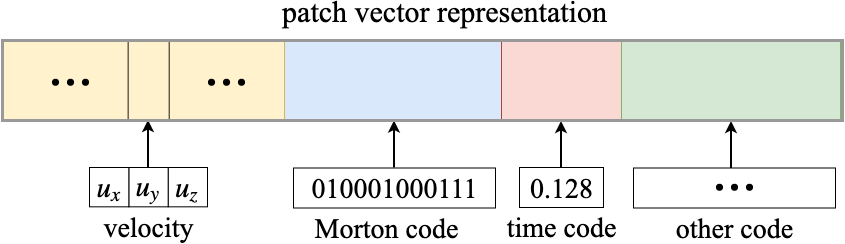}\vspace*{-3mm}
	\caption{\textbf{Augmented patch through space-time hard-coding.} We can use an augmented patch representation to improve spatial and temporal coherence: in addition to the velocity field, we add Morton code of the patch center, the time code of simulation time step the patch comes from, as well as any relevant other code involved in the simulation, such as the inlet size and position, to form our new patch representation vector.\vspace*{-2mm}}
	\label{fig:patch_representation}
\end{figure}

\paragraph{Space-time encoding}
For very contrived examples where there is no significant changes in the scene to upsample compared to the learning examples, we can add space and time encoding to the patch by augmenting each input patch vector with spatial and temporal components as sketched in Fig~\ref{fig:patch_representation}.
To encode the patch position, Morton codes~\cite{Karras-2012} can be used as they have nice locality properties compared to a simple 3D offset vector. 
For each local patch, the Morton code corresponding to its center is simply added to the representation vector of that patch.
For temporal encoding, the time step normalized by the maximum number of time steps of the simulation sequence can also be added to the representation vector.
In addition, to support variation of flow conditions, the various simulation parameters (such as different inlet sizes and positions) can be taken as extra codes to be added to the representation vector.
Knowledge of the position and time as well as the system parameters that a patch in the training set is coming from obviously guide the synthesis tremendously, as their relevance to a similar simulation is directly encoded into the patch representation. 
However, this brute-force encoding is \emph{very} rigid, and should only be employed for scenarios where the animation sequence  to be upsampled and simulation parameters are quite similar to the training simulations. 
Figs.~\ref{fig:overview_compare}(f), \ref{fig:re-simulation1} \& \ref{fig:re-simulation2} show how exceedingly well the results of this approach can perform, providing a very efficient exploration through learning of motions near a given set of animation sequences. \\[-5mm]

\paragraph{Phase-space encoding}
However, space-time encoding prevents more universal generality: if a simulation sequence to be upsampled is markedly different from any of the training simulation sequences, adding space and time information to the patches can in fact degrade the results as it implicitly guides the network to use patches in similar places and at similar times even if it is absolutely not appropriate in the new simulation. 
Instead, we wish to augment patch data with more information about the \emph{local dynamical behavior of the flow}. 
One simple idea is to use phase space information: instead of using only the local vector field stored as $\mathbf{y}_l$, we can encode the patch with the time history of this local patch: $[\mathbf{y}_l^{t},\mathbf{y}_l^{t-1}, ..., \mathbf{y}_l^{t-\tau}]$ where $\tau$ is the maximum number of previous time steps to use. 
Note that just picking $\tau\!=\!2$ corresponds in fact to the typical input of a full-blown integrator: knowing both the current and previous local vector fields is enough to know both velocity and acceleration of the flow locally. 
Our upsampling approach using this $\tau\!=\!2$ case can thus really be understood as a learned predictor-corrector integrator of the fine motion based on the two previous coarse motions: the coarse simulation serves, once directly upsampled to a higher-resolution grid, as a prediction, to which a correction is added via learned dynamic behaviors from coarse-fine animation pairs.
Figs.~\ref{fig:teaser}(a) \& \ref{fig:generalized_synthesis} show the results of such a phase-space representation, with which a variety of synthesis results can be obtained.

Comparing the results of the new patch encoding with the one containing only the velocity field in Fig.~\ref{fig:overview_compare}(e), we see that the augmented representation captures much improved coherent vortical structures without obvious noise.
While the synthesis results using a phase-space encoding may be slightly worse than the space-time encoded ones in terms of capturing the small-scale vortical structures of the corresponding high-resolution simulations, this significantly more general encoding can handle much larger differences (such as translations of inlets, rotations of obstacles, or even longer simulations than the training examples) in animation inputs.\\[-5mm]

\paragraph{Vorticity} For flows in general and smoke in particular, the visual saliency of vorticity is well known. Unsurprisingly, we found beneficial to also add the local vorticity field to the patch encoding: while this field is technically just a linear operator applied to the vector field, providing this extra information led to improved visual results, without hurting the learning rate. Consequently, and except for the few figures where space-time encoding is demonstrated, we always use only the last three vector fields and last three vorticity fields as the patch encoding, i.e., $[\mathbf{y}_l^{t}, \mathbf{y}_l^{t-1},\mathbf{y}_l^{t-2}, \nabla\!\times\!\mathbf{y}_l^{t}, \nabla\!\times\!\mathbf{y}_l^{t-1}, \nabla\!\times\!\mathbf{y}_l^{t-2}]$. \\[-2mm]

\begin{figure}[t]
	\centering
	\includegraphics[width=1.0\columnwidth]{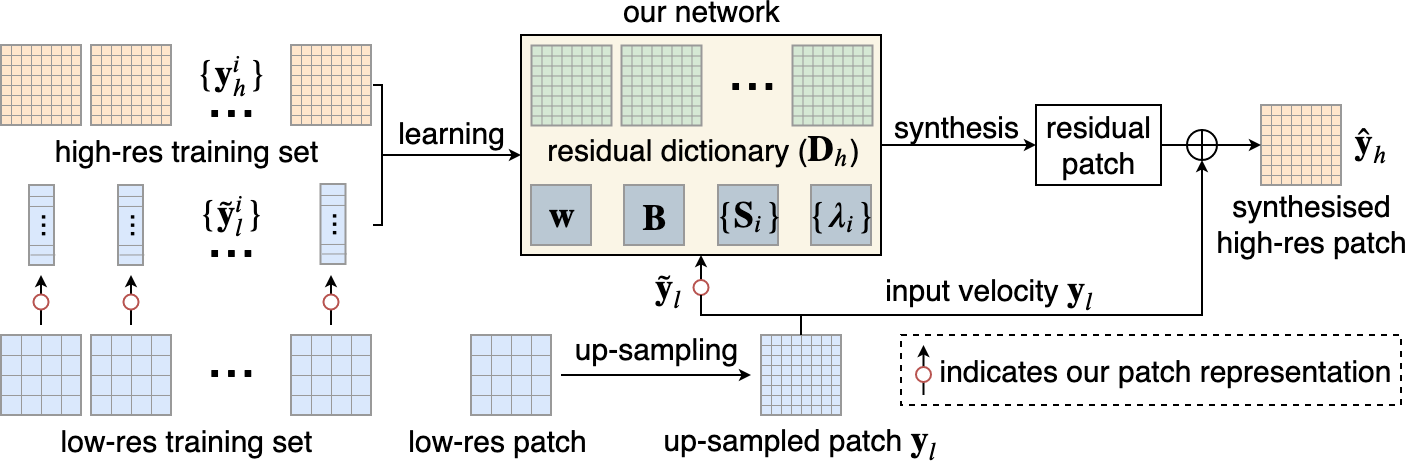}\vspace*{-2mm}
	\caption{\textbf{Our network-based dictionary learning approach.} In order to synthesize high-resolution flow fields, we first prepare a training set of local patch pairs ($\mathbf{y}_l^i$ and $\mathbf{y}_h^i$) from low- and high-resolution flow simulations respectively (left); note that the low resolution patches are represented by our augmented patch vector. With this training data, we learn a residual dictionary $\mathbf{D}_h$ as well as its associated predictor $\mathbf{w}_T(\mathbf{y}_l)$.  Given a low resolution flow field, each local patch is fed to the network to predict a set of sparse coefficient $\mathbf{w}$ such that the high resolution patch can be synthesized using $\mathbf{D}_h$ and $\mathbf{w}$ added to the upsampled input patch. \vspace*{-5mm}}
	\label{fig:our_learning_diagram}
\end{figure}

\paragraph{Rotation.}
When synthesizing general flows, the overall flow field may be rotated compared to the training examples, e.g., when a coarse flow with an inlet is rotated by 90 degrees or when an obstacle is rotated by 45 degrees.
In such a case, training from a set without this rotation may not lead to accurate results due to a lack of smoke motion in the proper direction (remember that we synthesize velocity fields, rather than density fields).
To tackle this problem, we simply add rotated versions of each local patch to the training set. 
Several rotation angles can be sampled, for instance, each $\pi/2$ rotation for each coordinate direction.
Fig.~\ref{fig:generalized_synthesis} shows a result using such a phase-space patch encoding including $\pi/2$-rotations, with coarse simulations containing obstacles that are rotated by 45 degrees in (e) \& (f) along different coordinate directions. Fig.~\ref{fig:our_learning_diagram} summarizes the overall workflow for synthesizing high-resolution flow fields with our new network and augmented patch encoding.

%\vspace{-0.2cm}
\subsection{Network learning}

The network we just described can be classically trained by providing a large number of training pairs of coarse and fine simulation patches (we will discuss how to judiciously select candidate patches from a set of coarse and fine animation pairs in a later section): the loss function $L_{T}(\Theta)$ has to be minimized with respect to all the parameters stored in $\Theta$, for instance, by the ``Adam'' method~\cite{Kingma-2014}, with full-parameter update during optimization.
However, a large number of layers and parameters may not produce good training convergence in case of large variety of motions in the training set. 
In order to improve convergence -- and thus, induce better prediction results during synthesis -- we can also employ a \emph{progressive learning algorithm} similar to~\cite{Borgerding-2017} which performs learning optimization in a cascading way, using the learned result from the previous layer as part of the initialization for the learning of the next layer.
As the learning of one single layer involves only a small fully-connected network, and because this cascading approach to learning gradually provides better initialization than the traditional full-parameter learning, the learning process turns out to exhibit better convergence.\\[-2mm]

\begin{figure}[t]
	\centering
	\includegraphics[width=\columnwidth]{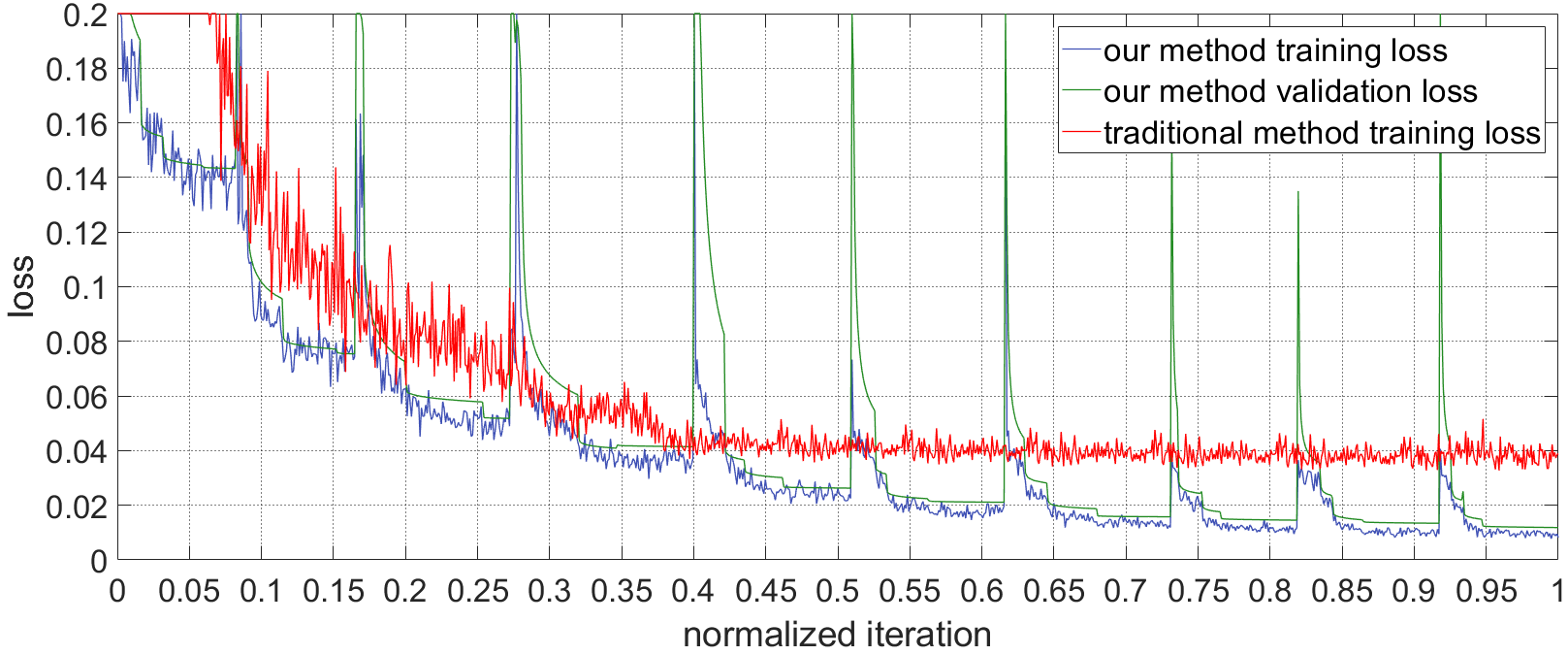}\vspace{-3mm}
	\caption{\textbf{Training convergence.} Our progressive training exhibits better convergence compared to direct, full-parameter learning.\vspace{-3mm}}
	\label{fig:loss}
\end{figure}

\begin{algorithm}[b]
	\caption{Pseudo-code of our progressive learning algorithm.}
	Set up a parameter set $\Theta$ with $\mathbf{D}_h$, $\mathbf{B}$, $\mathbf{S}_1$, $\lambda_1$ and $\lambda_{T+1}$.\\
	Initialize $\Theta$ with random numbers. \\
	\vspace{0.1cm}
	For ($i$= 1; $i$<$T$; $i$++) \hspace{0.2cm}// $T$ is the maximum number of layers \\
	\hspace{0.2cm} Learn $\Theta$ for layer $i$ to obtain $\mathbf{D}_h$, $\mathbf{B}$, $\lambda_{T+1}$, $\{\mathbf{S}_j\}$ and $\{\lambda_j\}$, $j=1,...,i$ .\\
	\hspace{0.2cm} Add $\mathbf{S}_{i+1}$ and $\lambda_{i+1}$ into the parameter set $\Theta$ . \\
	\hspace{0.2cm} Initialize $\mathbf{S}_{i+1}$ and $\lambda_{i+1}$ with random numbers . \\
	End For\\
	\vspace{0.1cm}
	Output learned parameters $\mathbf{D}_h$, $\mathbf{B}$, $\lambda_{T+1}$, $\{\mathbf{S}_i\}$ and $\{\lambda_i\}$, $i=1,...,T$.
	\label{alg:learning_algorithm}
\end{algorithm}

More specifically, we first initialize all variables randomly in $[-0.01,0.01]$ and perform learning for the first layer to find the optimal parameters $\smash{\mathbf{D}_h}$, $\smash{\mathbf{B}}$, $\smash{\mathbf{S}_1}$ and $\smash{\lambda_1}$.
We then use these parameters as initialization for the learning phase of the second layer, where now another set of parameters $\mathbf{S}_2$ and $\lambda_2$ are added (with random initial values), and this new learning results in another set of optimal parameters for all the variables involved.
This process repeats by adding $\mathbf{S}_{i+1}$ and $\lambda_{i+1}$ into the learning for the $(i\!+\!1)$-th layer, with all other parameters from previous layers initialized to the learning result of the $i$-the layer, until all the layers in the network are learned. 
For each learning phase, we also employ the ``Adam'' method~\cite{Kingma-2014}.
When using space-time encoding, we use $90\%$ of the training patches for learning and the remaining $10\%$ for validation;
when phase-space encoding is used, we found preferable to use training patches from several simulation examples, and use patches from different simulation example for validation to better test the generalization properties of the training.	
We obtain the final learning result once we reached the $T$-th (final) layer of the network.
Alg.~\ref{alg:learning_algorithm} illustrates the pseudo-code for our progressive learning process with better convergence properties. \\[-2mm]

\begin{figure}[t]
	\centering
	\includegraphics[width=0.97\columnwidth]{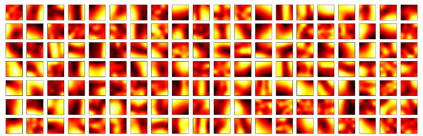}\vspace{-3mm}
	\caption{\textbf{Example of learned dictionary.} Visualization of cross-sections of 3D velocity patches from a portion of the dictionary set.\vspace{-3mm}}
	\label{fig:dictionary}
\end{figure}
We show in Fig.~\ref{fig:loss} the training loss (blue) as compared to the traditional training loss using full parameters (red), as well as the corresponding validation loss (green) during a typical training of our network. 
The periodic large peaks indicate a transition from one layer to the next, where a subset of the randomly initialized parameters are inserted into the learning, increasing the loss to a very high value; however, the loss quickly goes back down to an even smaller value. 
Compared to full-parameter learning, our progressive learning method converges to a smaller loss for both training and validation sets, thus enabling better synthesis results.\\[-2mm]

At the end of either full-parameter or progressive learning, we obtain all network parameters, including the dictionaries.
Fig.~\ref{fig:dictionary} shows a partial visualization of the learned dictionaries through small cross-sections of selected patches. %\vspace*{-1mm}
It should be noted that, although progressive learning can produce better convergence (and thus better synthesis results), it can also be much slower than full-parameter learning for large training sets. 
In practice, we compromise between learning accuracy and efficiency: we use full-parameter learning for cases where the diversity of the training set is relatively small, and progressive learning otherwise (see Tab.~\ref{tab:param}).

\begin{figure}[b]
	\centering
	\includegraphics[width=\columnwidth]{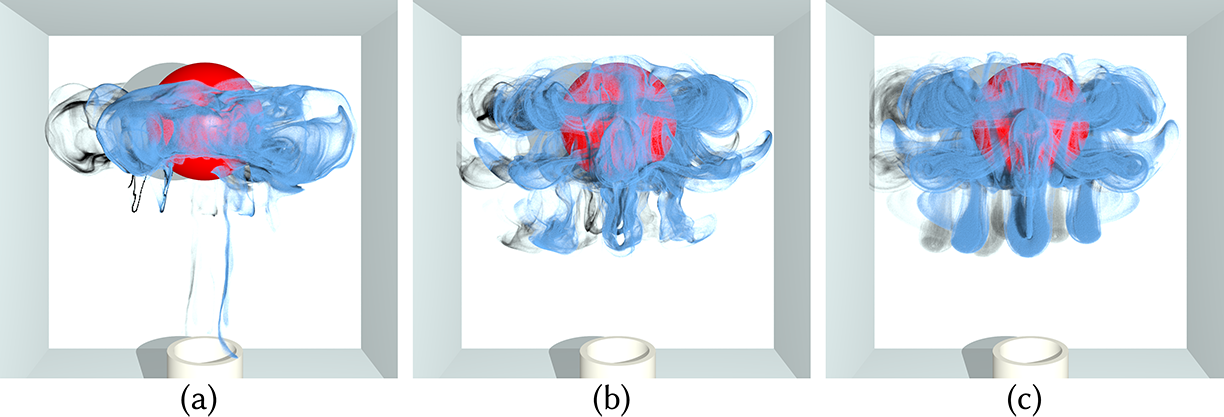}\vspace{-4mm}
	\caption{\textbf{Original \emph{vs.} multiscale synthesis.} From training simulations only containing one inlet on the left of the domain, simulating a bottom inlet produces an adequate, but inaccurate upsampling (a) due to limited rotation samples; the same simulation using our multiscale network (b) produces a result much closer to the corresponding fine simulation (c). \vspace{-1.5mm}}
	\label{fig:multiscale_comparison}
\end{figure}

\begin{figure*}[t]
	\centering
	\includegraphics[width=0.96\textwidth]{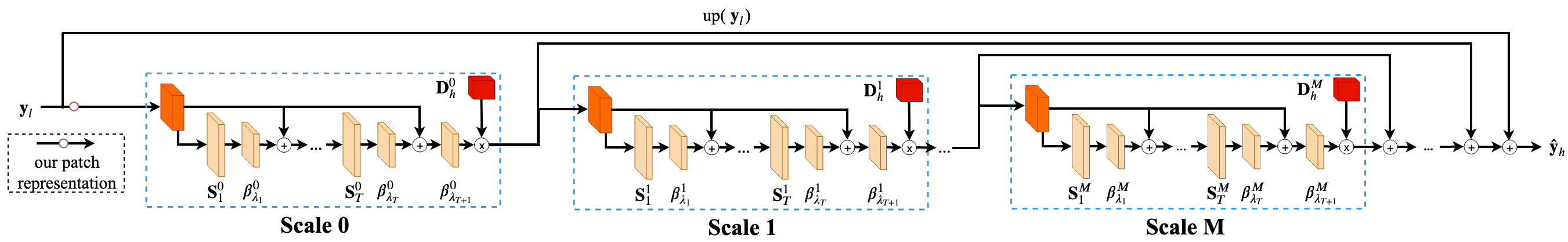}\vspace*{-4mm}
	\caption{\textbf{Multiscale network.} To increase the network representability, a multiscale version of our network can be employed. This network structure subdivides the residual patch into $M$ multiple scales, and each scale is represented and learned by our original network. The synthesis result is obtained by summing together all the components that each of these subnetworks synthesizes.\vspace*{-2mm}}
	\label{fig:multi-scale-network}
\end{figure*}

\subsection{Multiscale network}
If our training set has very high diversity, the design of our network described so far may no longer be appropriate as shown in Fig.\ref{fig:multiscale_comparison}(a) when rotated patches are added for training: if the training set contains too diverse a set of physical behaviors, $\mathbf{D}_h$ becomes too complex and can \emph{exceed} the representability of the network.
We could further increase the depth of the network and the size of the dictionary to increase the network capacity to handle more complex representations, but at the expense of significantly increased training and synthesis times. 
Instead, motivated by multi-resolution analysis, we decompose $\mathbf{D}_h$ into multiple scales (components): $\mathbf{D}_h = \mathbf{D}_h^0 + \mathbf{D}_h^1 + ... + \mathbf{D}_h^M$, where each scale is represented by our previous LISTA-like network, resulting in a multiscale network as depicted in Fig.~\ref{fig:multi-scale-network}. Even if each sub-network is rather simple in its number of layers and dictionary size (and thus limited in its complexity of representation), the cumulative complexity of the resulting $M$-scale network is significantly increased.
While the learning phase of this multiscale network could still follow the same progressive optimization process as we described above, we found it very slow to converge compared to a full-parameter optimization, for only a marginal gain in final loss. 
Thus, all of our examples based on a multiscale network (i.e., Figs.~\ref{fig:teaser}(a)\&(b),~\ref{fig:multiscale_comparison},~\ref{fig:generalized_synthesis} and~\ref{fig:restricted synthesis}) were trained via full-parameter optimization, with $M\!=2$ since a two-level hierarchy proved sufficient in practice.
Fig.~\ref{fig:multiscale_comparison}(b) shows the synthesis from such a multiscale network when rotated patches are added to the training set, indicating that much better results can be obtained with this multiscale extension when compared to the corresponding fine physical simulation shown in Fig.\ref{fig:multiscale_comparison}(c).
\subsection{Assembly of training set}
For network training, we need to prepare a large number of training pairs of corresponding low-resolution and high-resolution patches. \\[-3mm]

The patch size should be carefully chosen to tune efficiency and visual coherence.
Too small a size may not capture sufficient structures, whereas too large a size may require a very large dictionary and thus slower training and more non-zero coefficients during synthesis, hampering the overall computational efficiency. In practice, we found that a low-resolution patch size of $n\!=\!5$ (and whichever high-resolution size this corresponds to depending on the amount of upsampling targeted) is a good compromise, and \emph{all our results were generated with this patch size}. \\[-3mm]

\begin{figure}[t]
	\centering
	\includegraphics[width=0.96\columnwidth]{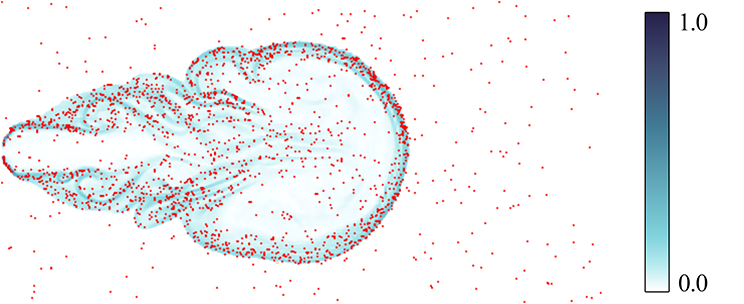}\vspace*{-2mm}
	\caption{\textbf{Importance sampling for training.} We select our training patches based on an importance sampling calculated from smoke density and local strain, where darker colors indicate higher importance (hence more selected patches); red dots show selected training patch centers.\vspace{-2.5mm}}
	\label{fig:sampling}
\end{figure}

In general, these small patches should come from a set of different simulation sequences with different boundary conditions, obstacles, or physical parameters to offer enough diversity for our training approach to learn from.
The training patch pairs are then automatically selected from these sequences. Instead of using the entire set of local patches from all sequences, we found that proper patch sub-selection is important for efficiency: getting a good diversity of patches gives better guidance for the network training and higher convergence, thus producing better synthesis results. 
We thus adopt an importance sampling strategy~\cite{Kim-2013}, where a Poisson-disk driven selection of patches is done with a probability distribution based on the density of smoke (i.e., on the local number of passive tracers advected in the flow to visualize the smoke) on either low- or high-resolution simulations: in essence, we favor the regions where smoke is likely to be or to accumulate during an animation to better learn what is visually most relevant.
Another criterion of visual importance that we found interesting to leverage during patch selection is a large local strain rate: since turbulent flows are particularly interesting due to their small scale structures, targeting predominantly these regions where wisps of smoke are likely to be present allows the network to better synthesize these salient features.
Fig.~\ref{fig:sampling} shows an illustration of such an importance sampling, where color luminosity indicates sampling importance.

\begin{table*}
	\small
	\centering 
	\begin{center}
		\begin{tabular}{cccccccc}
			\toprule
			Items&Fig.~\ref{fig:teaser}(a) \& Fig.~\ref{fig:generalized_synthesis}&Fig.~\ref{fig:teaser}(b) \& Fig.~\ref{fig:restricted synthesis}&Fig.~\ref{fig:teaser}(c) \& Fig.~\ref{fig:re-simulation2}&Fig.~\ref{fig:re-simulation1} & Fig.~\ref{fig:diff_res}(d)& Fig.~\ref{fig:diff_res}(e) & Fig.~\ref{fig:diff_res}(f)\\
			\midrule
			Resolution (coarse)               &50$\times$50$\times$50 &50$\times$50$\times$50 &100$\times$50$\times$100 &30$\times$90$\times$90 &25$\times$37$\times$25 &25$\times$37$\times$25 &25$\times$37$\times$25\\
			Resolution (fine)             &200$\times$200$\times$200 &200$\times$200$\times$200 &400$\times$200$\times$400 &120$\times$360$\times$360 &50$\times$75$\times$50 &100$\times$150$\times$100 &200$\times$300$\times$200\\
			Network structure     &multiscale &multiscale &single scale &single scale &single scale &single scale &single scale\\
			Learning method         &full-parameter &full-parameter &progressive &progressive &progressive &progressive &progressive\\
			Patch encoding method       &phase-space &phase-space &space-time &space-time &space-time &space-time &space-time\\
			Dictionary size       &800 &800 &800 &800 &400 &400 &400\\
			Network memory size        &49M &49M &22M &22M &4M &8M &9M\\
			Training setup time  &16 hours &20 hours &8 hours &6 hours &2 hours &3 hours &5 hours\\
			Training time         &72 hours &65 hours &37 hours &31 hours &15 hours &19 hours &27 hours\\
			Time cost (coarse)         &0.06 sec. &0.06 sec. &0.12 sec. &0.09 sec. &0.02 sec. &0.02 sec. &0.02 sec.\\
			Time cost (fine)           &9.7 sec. &9.7 sec. &75.9 sec. &20.1 sec. &0.07 sec. &0.89 sec. &11.2 sec.\\
			Time (upsampling)     &2.9 sec. &2.9 sec. &3.48 sec. &1.75 sec. &0.04 sec. &0.12 sec. &0.77 sec.\\
			Speed-up              &3.3 &3.3 &21.8 &11.5 &1.8 &7.4 &14.5\\	
			\bottomrule
		\end{tabular}
	\end{center}
	\caption{\textbf{Statistics.} We provide in this table the parameters, timings and memory use per frame for various smoke animations shown in this paper.} 
	\vspace{-6mm}
	\label{tab:param}
\end{table*}

\subsection{High-resolution flow synthesis}
\label{sec:turbulence_detail_synthesis}

\begin{figure}[t]
	\centering
	\includegraphics[width=0.75\columnwidth]{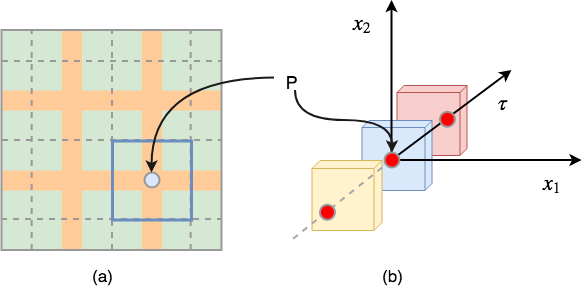}\vspace*{-3mm}
	\caption{\textbf{Patch blending.} 2D illustration of 4D convolution for overlapped patches: (a) patches are laid out with overlapping, where overlapped regions are in orange; (b) for each node in the overlapped region, the overlapping space is shown along the $\tau$ direction, where the center of coordinate in that space is placed at the patch with no overlap, see the dotted line in (a).\vspace*{-1mm}}
	\label{fig:boundary_treatment}
\end{figure}

After learning, the network automatically predicts high-resolution patches from low-resolution input ones by evaluating the local sparse weights that best reconstruct the dynamical surrounding of each patch. To further improve spatial coherency, we evaluate overlapping high-resolution patches, then blending of the overlapped regions (see orange regions in Fig.~\ref{fig:boundary_treatment}(a) as an example) is performed (in parallel) to ensure a smooth global reconstruction. 
Different blending approaches could be used; we settled on a convolution-based method as follows.
We consider the synthesized velocity $\mathbf{u}(\mathbf{x},\tau)$ in overlapped regions as a 4D function separately in 3D space ($\mathbf{x}$) and an overlapping space coordinate ($\tau$) (see Fig.~\ref{fig:boundary_treatment}(b)), and employ a 4D Gaussian kernel $\mathbf{G}(\sigma_\mathbf{x}, \sigma_\mathbf\tau)$ to do the convolution, with $\sigma_\mathbf{x}$ and $\sigma_\mathbf\tau$ the standard deviations for spatial and overlapping domains respectively. 
We set $\sigma_\mathbf{x}\!=\!2.5$ and $\sigma_\mathbf\tau\!=\!1.5$ in all our experiments.
Since the whole convolution is separable, it can be formulated as:\vspace*{-1.5mm}
\begin{equation}
\mathbf{u}(\mathbf{x},\tau) \leftarrow \mathbf{G}(
\sigma_\tau) \ast \left[\mathbf{G}(\sigma_\mathbf{x}) \ast \mathbf{u}(\mathbf{x},\tau)\right].\vspace*{-1.5mm}
\end{equation}
This means that we first conduct a 3D convolution in the spatial domain followed by a 1D convolution in overlapping space after local patch prediction to obtain the final synthesized result for the whole high-resolution field.
%Note that for convolution in overlapping domain, the patch with uniform non-overlapped subdivision is taken as the center of the Gaussian kernel $\mathbf{G}(\sigma_\tau)$, see Fig.~\ref{fig:boundary_treatment} for an illustration.
%\input{6.turbulence-detail-synthesis.tex}
%\begin{figure*}[t]
%	\centering
%	\includegraphics[width=\textwidth]{images/result1.png}
%	\caption{Simulation of vortex ring smoke injected from an inlet in the left of the domain. (a) low resolution simulation input ($50\times100\times50$); (b) our synthesized smoke at high resolution ($200\times400\times200$); (c) high resolution simulation with the same resolution as (b) for reference. It is obvious that important visual structures such as the two rotating vortices can be well synthesized that resemble the corresponding physical simulation.
%	}
%	\label{fig:result1}
%\end{figure*}

\begin{figure*}[t]
	\centering
	\includegraphics[width=0.98\textwidth]{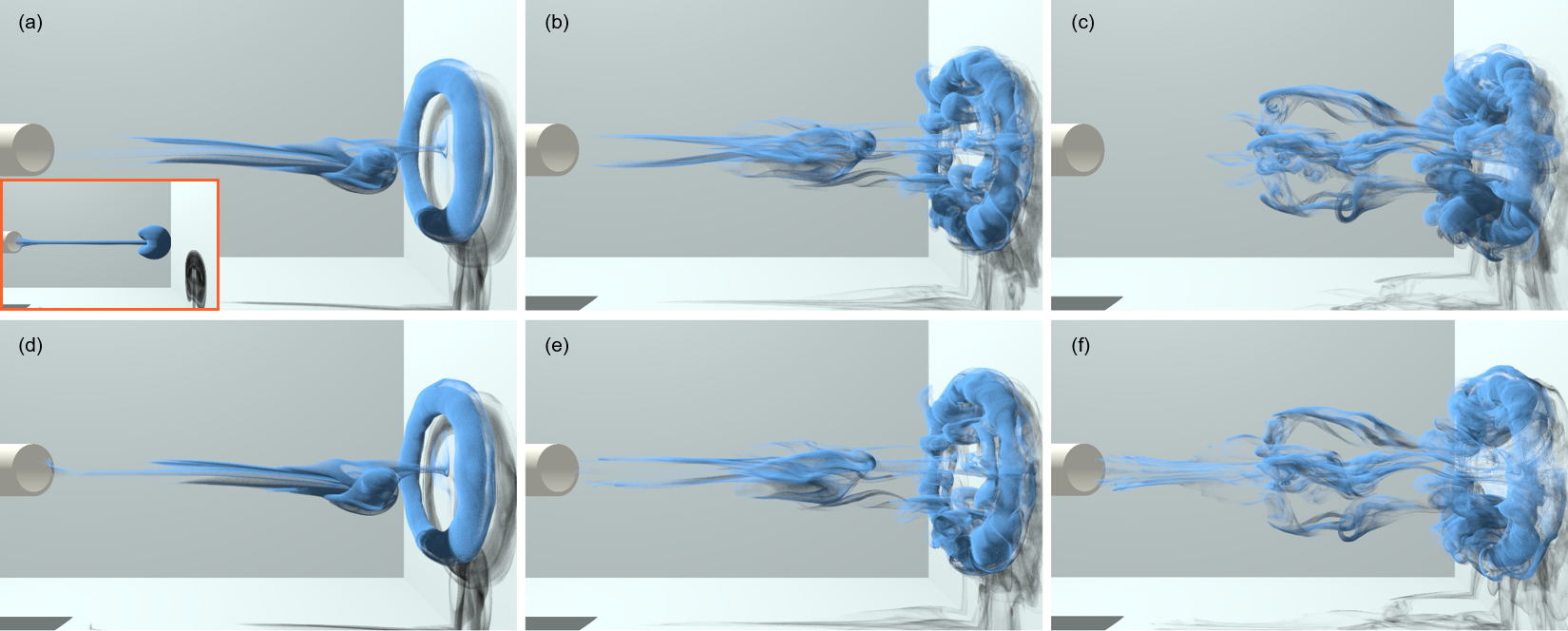}\vspace*{-2mm}
	\caption{\textbf{Different upsampling factors.} From the low resolution ($25\times37\times25$) smoke flow shown as an inset, the corresponding fine (top) and synthesized (bottom) animations are shown at different resolutions: (a/d) $50\times75\times50$, (b/e) $100\times150\times100$ and (c/f) $200\times300\times200$. \vspace*{-1.0mm}}
	\label{fig:diff_res}
\end{figure*}

%\begin{table*}
%	\centering 
%	\begin{center}
%		\begin{tabular}{cccccccc}
%			\toprule
%			Figure&Low-Res&High-Res&Time (Coarse)&Time (Fine) & Time (Upsampling)& Speed-up & Training Time\\
%			\midrule
%			Fig.~\ref{fig:teaser} (a) \& Fig.~\ref{fig:generalized_synthesis} &50$\times$50$\times$50 &200$\times$200$\times$200 &0.06 sec. &9.7 sec. &2.9 sec. &3.3 &120 hours\\
%			Fig.~\ref{fig:teaser} (b) \& Fig.~\ref{fig:restricted synthesis} &50$\times$50$\times$50 &200$\times$200$\times$200 &0.06 sec. &9.7 sec. &2.9 sec. &3.3 &65 hours\\
%			Fig.~\ref{fig:teaser} (c) \& Fig.~\ref{fig:re-simulation2} &100$\times$50$\times$100 &400$\times$200$\times$400 &0.12 sec. &75.9 sec. &6.9 sec. &11.0 &37 hours\\
%			Fig.~\ref{fig:re-simulation1} &30$\times$90$\times$90 &120$\times$360$\times$360 &0.09 sec. &20.1 sec. &3.5 sec. &5.7 &31 hours\\
%			%Fig.11(c)&300$\times$600$\times$300 &2.5min/f & $2\times10^{-3}$ & $2\times10^{-2}$\\		
%			\bottomrule
%		\end{tabular}
%	\end{center}
%	\caption{\textbf{Statistics.} We provide in this table the parameters and timings/\XP{memory} per frame of animation for various smoke animations shown in this paper.} 
%	\vspace{-6mm}
%	\label{tab:param}
%\end{table*}

\begin{figure}[t]
	\centering
	\includegraphics[width=0.97\columnwidth]{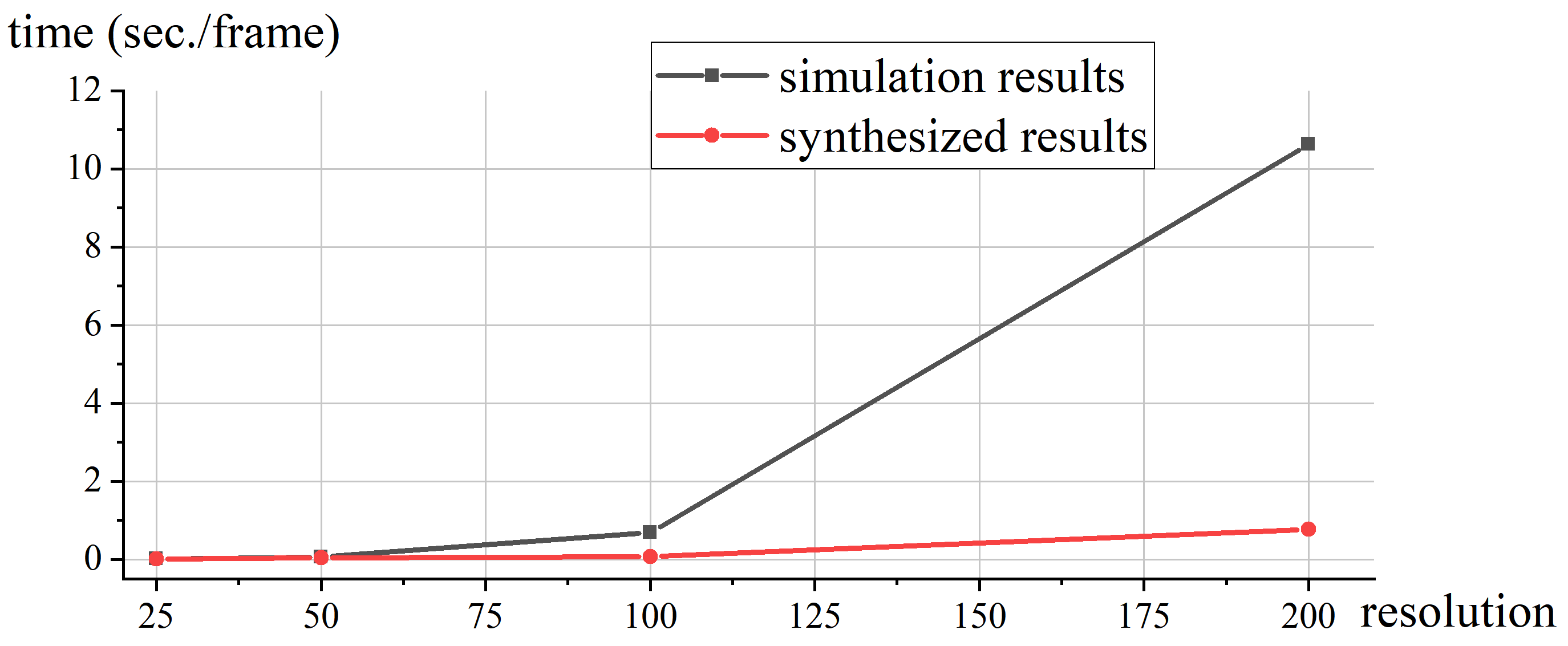}\vspace*{-3mm}
	\caption{\textbf{Speedup.} Comparison of performance under different resolutions for the upsampling example in Fig.~\ref{fig:diff_res} with restricted synthesis; gray curve indicates numerical simulation time at different resolutions, while red curve displays coarse simulation plus synthesis time for the same resolutions.\vspace*{-4mm}}
	\label{fig:performance}
\end{figure}

\section{Implementation Details}
\label{sec:implementation_details}

While our approach can handle basically any dictionary or patch size, we discuss the various choices of implementation parameters we used in our examples next for reproducibility. 

\begin{figure*}[t]
	\centering
	\includegraphics[width=0.98\textwidth]{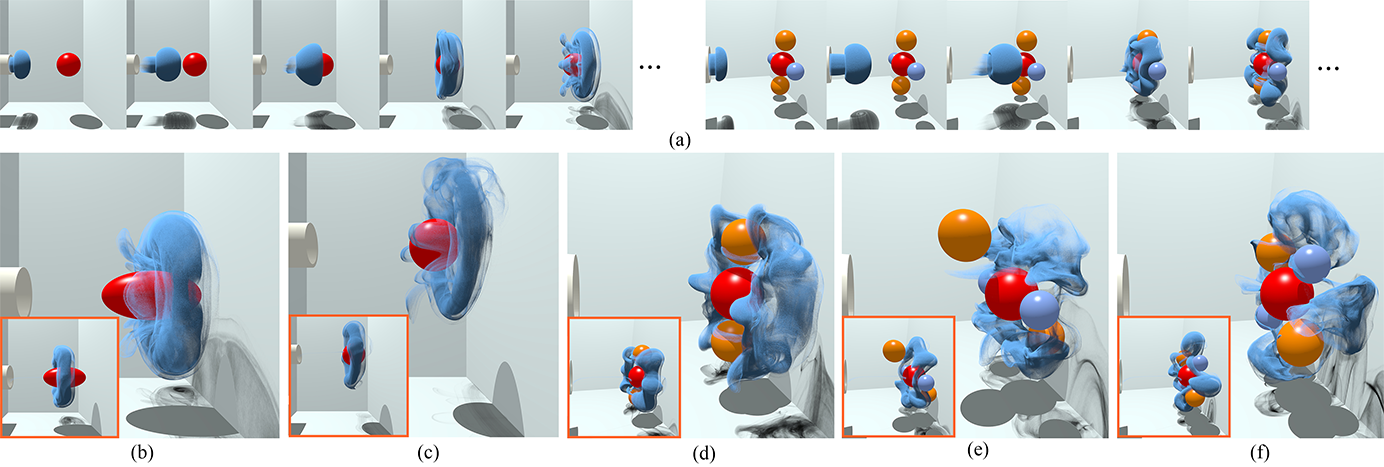}\vspace*{-4mm}
	\caption{\textbf{Generalized synthesis.} From two simulation examples containing different numbers of ball obstacles (a), our network-based approach can upsample (with ratio $64\!=\!4\!\times\!4\!\times\!4$) coarse simulations where we can change the shape of the ball (b), move the inlet position with longer simulation time (c), remove ball obstacles from the training set (d), and rotate the ball obstacles (e) \& (f), to show the generalizability of our network.\vspace*{-1mm}}
	\label{fig:generalized_synthesis}
\end{figure*}

\begin{figure*}[t]
	\centering
	\includegraphics[width=0.98\textwidth]{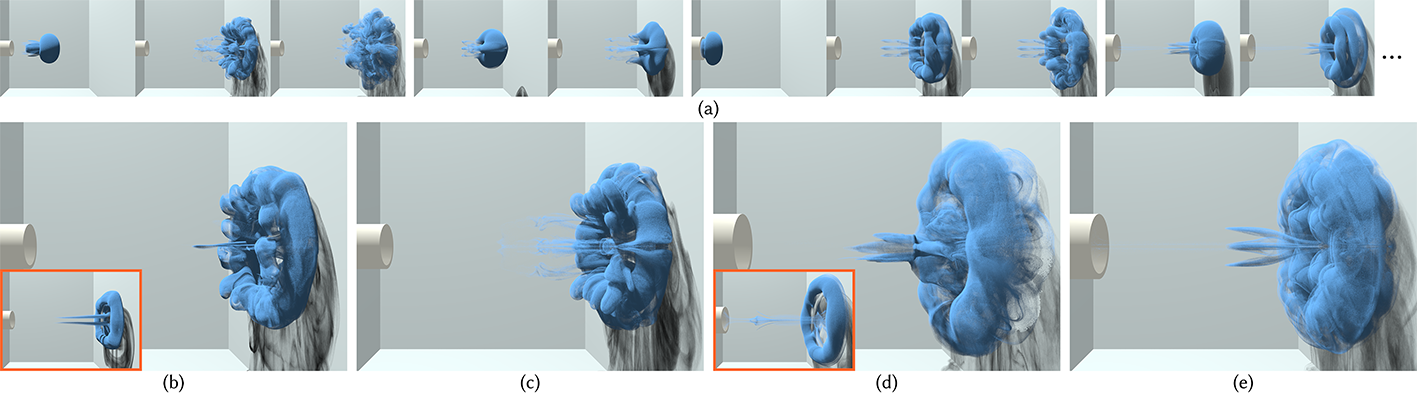}\vspace*{-4mm}
	\caption{\textbf{Restricted synthesis.} From a series of input coarse/fine animation sequences with only changes of the inlet size (a), our network-based approach can upsample smoke simulations (with ratio $64\!=\!4\!\times\!4\!\times\!4$) with arbitrary inlet sizes in between those used in the training set: (b) \& (d) for two different inlet sizes not present in the training set. Compared to the corresponding ground-truth numerical simulations (c) \& (e), our synthesized results share close resemblance.\vspace*{-2mm}}
	\label{fig:restricted synthesis}
\end{figure*}

\paragraph{Learning parameters.}
The dictionary size can be arbitrarily set, with larger size providing better results but slower training. 
We set it to 800 in the generalized synthesis case; other synthesis cases can be seen in Tab.~\ref{tab:param}.
As a rule of thumb, we recommend slightly larger values for higher Reynolds numbers --- and conversely, smaller values for lower Reynolds numbers --- to adapt to the complexity of the flow. 
For generalized synthesis with large variation of flow conditions compared to the training simulations (see, e.g., Fig.~\ref{fig:teaser}(a) \& Fig.~\ref{fig:generalized_synthesis}), we used 22M patches (selected via importance sampling from 800 frames across different simulations) to learn our dictionary and LISTA-like sparse coding layers; for more restricted synthesis cases where the variation of flow conditions is not too significant, the number of training patches can be lower: we used 15M for Fig.~\ref{fig:teaser}(b) \& Fig.~\ref{fig:restricted synthesis}; for re-simulation where there is only slight variation of the flow and obstacle positions, the number of training patches can be even lower: we used only 3M patches for Fig.~\ref{fig:teaser}(c) and Fig.~\ref{fig:re-simulation2}. 

\paragraph{Training details.}
In our implementation, we combine all patches that were collected for training to form a large matrix as input.
The learning process is then achieved by a series of matrix products, which are evaluated in parallel by CUDA with the CUBLAS library~\cite{Nvidia-2008}. 
During learning, since we need to compute the gradient tensor which is extremely large, we sample $4096$ patches for its computation.
The learning rate $h$ involved in the parameter matrix $\mathbf{B}$ is dynamically changed: initially, it is given a relatively large value, e.g., $h\!=\!0.0001$;
as iterations converge with this fixed $h$, we further decrease it until final convergence, i.e., when further reducing $h$ does not change the loss anymore.

\paragraph{Synthesis.}
The synthesis process is also implemented by a series of matrix products in parallel. We first collect all overlapped local patches to form a large matrix as input, and then go through the network by a series of parallel matrix calculations for synthesizing the high-resolution patches. 
A parallel convolution in overlapped regions is finally performed to obtain the synthesized high-resolution field, in which passive tracers can then be advected to render the smoke.
As for time discretization, we used a simple setup in which every simulation uses ten time steps between rendered frames to offer a good approximation of the flow to carry particles for smoke visualization. We then upsample every third coarse step to offer a good advection approximation with three upsampled vector fields between two high-resolution rendered frames.

%\begin{figure*}[t]
%	\centering
%	\includegraphics[width=\textwidth]{images/result3.png}
%	\caption{Simulation of a jet smoke flow passing a ball object. (a) low resolution simulation input ($50\times75\times50$); (b) our synthesized smoke at high resolution ($200\times300\times200$); (c) high resolution simulation with the same resolution as (b) for reference. It is obvious that important visual structures are  captured which are similar to the corresponding high-resolution simulation.}
%	\label{fig:result3}
%\end{figure*}

\paragraph{Libraries and memory requirements.}
Our learning was implemented with TensorFlow~\cite{Abadi-2016} on a server with NVIDIA P40 GPUs, each with a total memory of $24$GB.
For large training set (larger in size than the allowable GPU memory), we perform out-of-core computing by evaluating the loss and gradient with several passes and data loads.
The synthesis process was implemented on a basic workstation equipped with an NVIDIA TITAN X GPU with $12$GB memory.
The final rendering is achieved with a particle smoke renderer~\cite{Zhang-2015} together with the NVIDIA OptiX ray-tracing engine~\cite{Parker-2010}, which usually takes about $40$ seconds to render one frame of animation for a resolution of $1280\times720$ as the output image with multisample anti-aliasing ($3\times3$ samples per each output pixel), and with a maximum number of particles equal to $1.5\times10^7$. 
After learning, the whole network (including the resulting dictionary) takes up a total size of approximately 50MB for the generalized synthesis case (Fig.~\ref{fig:generalized_synthesis});
for the network size of other synthesis cases, refer to Table.~\ref{tab:param} for more specific details.

\paragraph{Timings and resulting compression.} 
As for timings, using the largest training set of patches in the generalized synthesis case (Fig.~\ref{fig:generalized_synthesis}) resulted in a $72$-hour training phase.
For more restricted cases, it takes an average of approximately $50$ hours to train the network.
In the very restricted re-simulation case, $15$ hours are typically needed for training.
During synthesis, it only takes between 1 to 3 seconds to synthesize one time step of a high resolution simulation ($200 \times 200 \times 200$) from a low resolution input ($50 \times 50 \times 50$).
The high-resolution flow synthesis after network learning is then much faster (often by an order of magnitude or more) than the corresponding physical simulation (see Fig.~\ref{fig:performance} for performance acceleration under different synthesis resolutions corresponding to Fig.~\ref{fig:diff_res}). 
In addition, memory requirement is also significantly reduced; e.g., a high-resolution simulation typically requires from $1600$MB to $3200$MB at the resolution of $200\times200\times 200$ (depending on the solver), while its low-resolution simulation only requires $25$MB to $50$MB at the resolution of $50\times50\times 50$, and in the generalized synthesis (Fig.~\ref{fig:generalized_synthesis}), the whole network only requires a total of $49$MB for synthesizing the high-resolution simulation.
Our network learning can thus be considered as a storage-efficient spatial and temporal compressor of fluid flows.
\section{Results and Discussions}
\label{sec:results_discussions}

%Figs.~\ref{fig:teaser},~\ref{fig:overview_compare},~\ref{fig:result1},~\ref{fig:result2} and~\ref{fig:result3} show our synthesized results with low resolution simulations as inputs, which are all compared with their high resolution simulation counterparts as ground-truths.
%Due to higher accuracy to preserve turbulence structures, we employ the recent kinetic method~\cite{Li-2018} to generate both low and high resolution velocity fields for learning.
%It can be noticed that many important structures in high resolution simulations do not exist in low resolution counterparts, which place a great challenge to synthesize high resolution fields that well approximate their corresponding simulations without any knowledge from them.
%By learning from high resolution simulations, the network can capture many important structures, such as vortices with relatively large sizes, and faithfully synthesize them from low resolution simulations, e.g., in Fig.~\ref{fig:result2}, a secondary vortex ring is well captured by our network (see the red box), which is visually important and physically studied in~\cite{New-2016}, indicating the strength of our method to preserve necessary important flow structures during synthesis.
%This is in stark contrast with existing methods for doing similar tasks.

We now discuss the various results presented in this paper. Most of the datasets used for training the network and synthesizing our results were collected from the recent kinetic fluid simulation method by~\cite{Li-2018}, due to its accurate simulation of turbulent flows.
However, our method is not restricted to a specific fluid solver: we can start from an arbitrary set of time-varying vector fields.

\subsection{Results in various scenarios}
%Dynamic upsampling of smoke flows can be applied towards a variety of goals as we now discuss.
We review our results according to the three different scenarios we described earlier, allowing for different generalization behaviors and synthesis accuracy.
\\[-4mm]

\paragraph{Generalized synthesis}
Arguably the most general approach for upsampling is to collect a large set of training patches from a variety of simulations covering very different flow conditions. 
We demonstrate in Fig.~\ref{fig:generalized_synthesis} that,
while this is resource intensive as it seemingly requires a good sampling of smoke behaviors for different parameters of the scene (here, inlet position and diameter, obstacle position, size and orientation, etc), even only two example simulations (Fig.~\ref{fig:generalized_synthesis}(a)) containing completely different configurations of ball obstacles is enough to train our dictionary-based upsampling process: the resulting network can handle very different coarse simulations, including changing the shape of the ball (Fig.~\ref{fig:generalized_synthesis}(b)), shifting the inlet position while increasing the simulation duration (Fig.~\ref{fig:generalized_synthesis}(c)), removing some of the ball obstacles
(Fig.~\ref{fig:generalized_synthesis}(d)), as well as rotating the ball obstacles by 45 degrees (Fig.~\ref{fig:generalized_synthesis}(e)\&(f); this configuration is not present in the training set since only 90-degree patch rotations are involved in the training set).
Although the synthesized high-resolution simulations are unlikely to match their corresponding fine physical simulations closely, the trained network captures complex and plausible vortex structures, far better than if only noise-like high-frequency structures were added to the coarse simulations. This illustrates the power of our approach: a few training simulations can already serve as a decent learning catalog to upsample a coarse simulation.\\[-4mm]

\begin{figure*}[t]
	\centering
	\includegraphics[width=\textwidth]{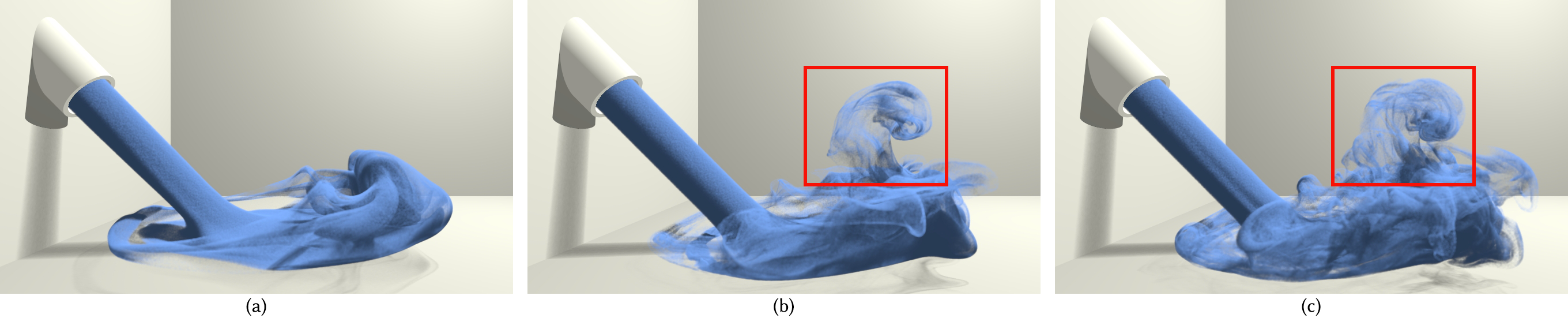}\vspace*{-4mm}
	\caption{\textbf{Smoke shooting.} From a low-resolution ($30\times90\times90$) simulation input (a), our synthesized smoke (b) at high resolution ($120\times360\times360$), vs. the high-resolution simulation (c) for reference. Despite a factor of 64 ($4\times4\times4$) in resolution ratio, visually important structures (e.g., the secondary vortex ring marked with a red box) which were not in the coarse simulation (a) but present in the fine simulation are well captured.\vspace*{-2mm}}
	\label{fig:re-simulation1}
\end{figure*}
\begin{figure*}[t]
	\centering
	\includegraphics[width=\textwidth]{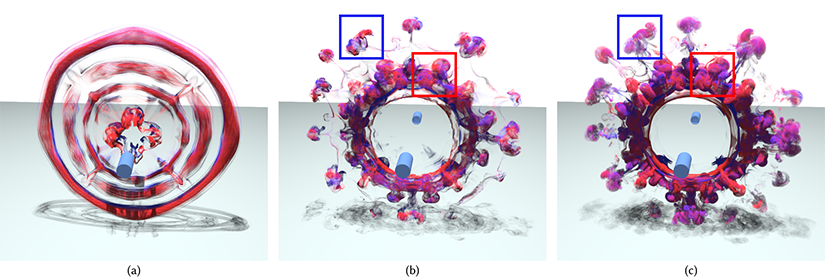}\vspace*{-4mm}
	\caption{\textbf{Vortex rings colliding.} From a low-resolution ($100\times50\times100$) simulation input (a), our synthesized smoke (b) at high resolution ($400\times200\times400$), vs. the high-resolution simulation (c) for reference. Despite a factor of 64 ($4\times4\times4$) in resolution ratio, we can still faithfully capture the obviously important vortex structures (e.g., the first (red box) and secondary (blue box) vortices, with leapfrogging in the center) present in the fine simulation.\vspace*{-1mm}}
	\label{fig:re-simulation2}
\end{figure*}

\paragraph{Restricted synthesis}
For better training and synthesis with the same or less computing resources as in the previous generalized synthesis case, we can reduce the variety of training simulations.
For example, for the jet-flow smoke shown in Fig.~\ref{fig:restricted synthesis}, we collect training patches from simulations using four different inlet sizes only, with phase-space encoding and with inlet size as an additional patch code, to synthesize high-resolution simulation results from a coarse simulation with an arbitrary inlet size in between those used in the training set.
Here, the training set shown in Fig.~\ref{fig:restricted synthesis} contains different smoke simulations with the largest inlet nearly twice as large as the smallest one, with two additional inlet sizes in between them, to produce a total of four simulation sequences, from which training patches are sampled.
Comparing to the high-resolution simulations as a reference (Figs.~\ref{fig:restricted synthesis}(c) \& (e)), the synthesized high-resolution flows contain vortex structures closely resembling the real, fine simulations.
Similar restricted cases, e.g., where we change the position or size of the obstacle in the flow, can be done as well.\\[-4mm]

\paragraph{Re-simulation}
In the extreme, we can use training patches sampled from a \emph{single} simulation, and the final synthesized high-resolution simulation can be considered as a re-simulation based on a quickly-evaluated coarse simulation for which the user has performed very small changes on the initial and/or boundary conditions, see Figs.~\ref{fig:re-simulation1} \&~\ref{fig:re-simulation2}.
This re-simulation case is much narrower in its applicability for upsampling, but can produce near-perfect synthesis results, achieving simulations that are very close in their vortical structures (see the secondary vortices in Figs.~\ref{fig:re-simulation1} \&~\ref{fig:re-simulation2}) to their corresponding physical simulations. 
The quality depends of course on the underlying Reynolds number, though: the lower the Reynolds number, the closer the synthesized simulation to its fine simulation counterpart.\\[-2mm]

\textit{Please check our supplementary video to see results for these three types of flow synthesis, with comparisons to their corresponding high-resolution simulations.}\\[-4mm]

\begin{figure}[t]\vspace*{-2mm}
	\centering
	\includegraphics[width=0.96\columnwidth]{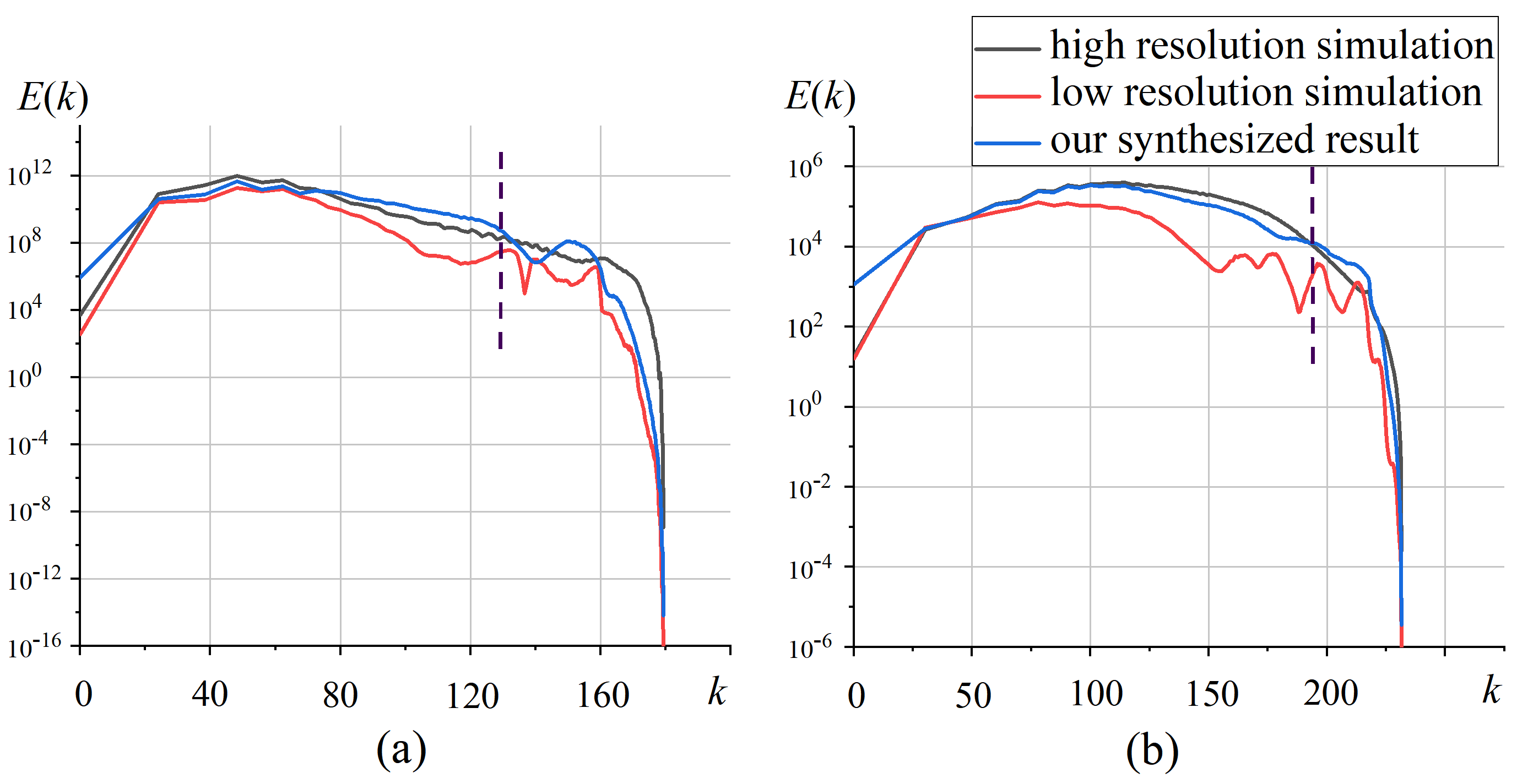}\vspace*{-4mm}
	\caption{\textbf{Spectral behavior.} We plot the energy spectra for a low-resolution simulation (red), a high-resolution simulation (gray) and our synthesized flow (blue) with respect to wavenumber $k$, and from two types of synthesis methods: (a) generalized synthesis, and (b) re-simulation. Although both the low-resolution energy spectra fluctuate significantly for large $k$, our network can faithfully produce spectra close to its high-resolution simulation counterparts below a critical wavenumbers (dotted lines).\vspace*{0mm}}
	\label{fig:energy_spectrum}
\end{figure}

\subsection{Synthesis accuracy}
As we highlighted early on, the technique proposed in this paper is not intended to generate high-resolution flow fields in close agreement to their physical simulations in all cases.
Since we target visually realistic smoke animations for relatively high Reynolds numbers, our method only ensures that faithful or plausible fine vortex structures present in the related physical simulations be generated from the coarse input, without noticeable artifacts.
There are several factors that affect the synthesized results.
The two main parameters are the dictionary size and the number of network layers, which both influence the dimensionality of the space of synthesized patches.
For flows with higher Reynolds numbers, one should use larger dictionary size since the local structures tend to be more complex, but larger dictionary size may lead to slower synthesis.
Another factor is how the training patches are sampled from the input coarse-fine animation pairs. 
In general, our method can capture most high-resolution flow structures, but our importance sampling may miss vortices that are only active for a very short period of time, and therefore, our synthesis will not capture them properly by lack of training. To a certain extent, the user may define a different notion of importance that highlights the most desirable features that synthesis is expected to recover. \\[-4mm]

\paragraph{Energy spectrum}
One of the important measures of accuracy, particularly for turbulent flows, is the energy spectrum.
Here, we show the spectral behavior in Fig.~\ref{fig:energy_spectrum} for the generalized synthesis (a) and re-simulation (b) cases.
Below a certain critical wavenumber (indicated via dotted lines), both spectra plots match the corresponding simulations well, indicating that both types of synthesis methods can retain large-scale vortex structures present in the high-resolution simulations, and re-simulation has a higher critical wavenumber, meaning that it can better capture small-scale vortex structures.\\[-4mm]

\begin{figure}[t]
	\centering
	\includegraphics[width=\columnwidth]{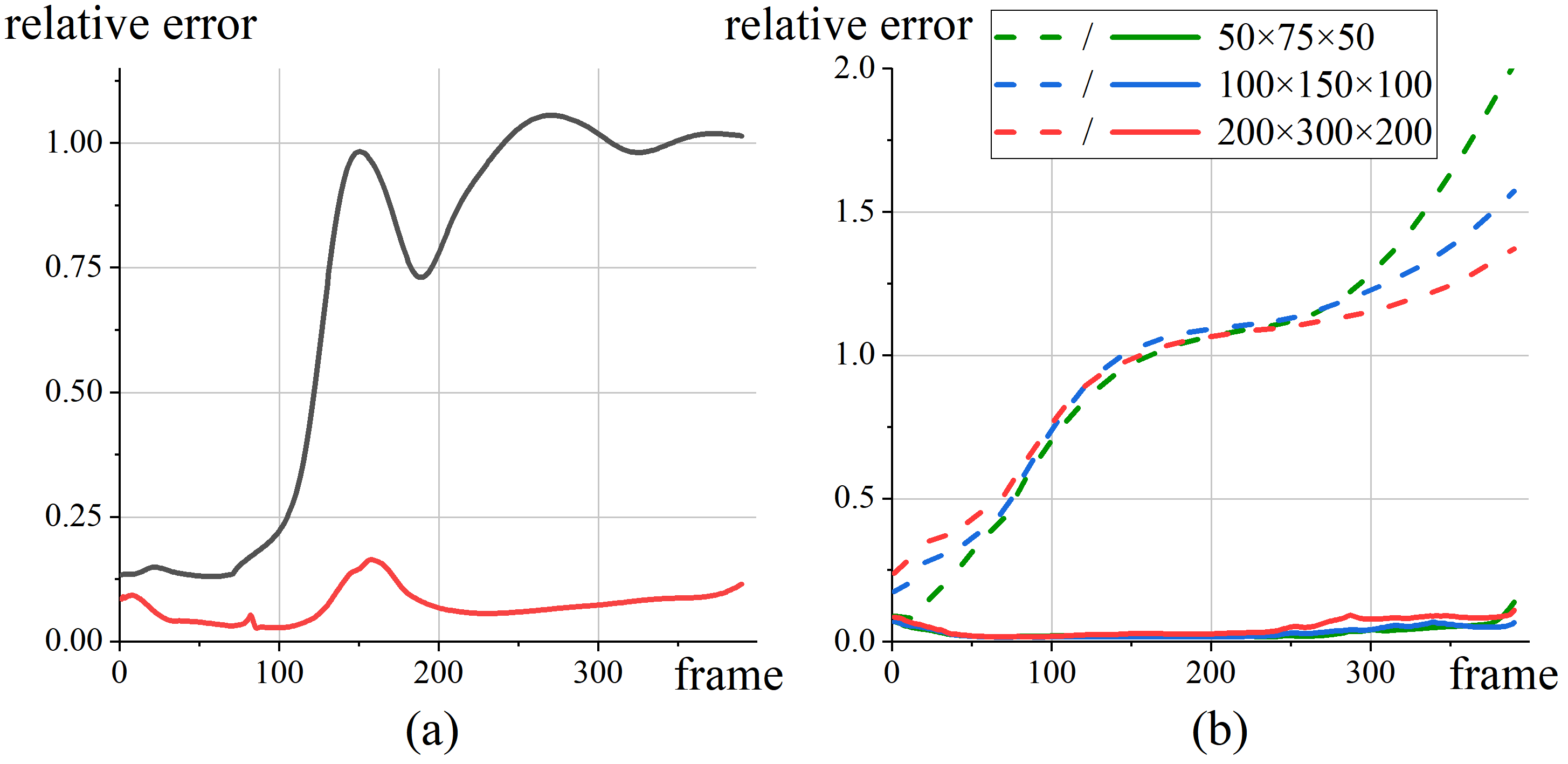}\vspace*{-4mm}
	\caption{\textbf{Synthesis error over time steps.} In (a), we plot relative errors for a generalized synthesis result (where the error is expected to be the largest), with a red curve for the error between our synthesized and high-resolution simulated flows, and a gray curve for the error between the coarse and fine simulations. In (b), we plot the error for upsampling the same low-resolution simulation input (25$\times$37$\times$25), this time to different high resolutions, where dotted lines indicate error between coarse and fine simulations as reference.\vspace*{-4mm}}
	\label{fig:relative_error}
\end{figure}

\paragraph{Synthesis error over time steps}
Another way to assess the accuracy of our synthesis result is to compute the mean squared error of velocity fields normalized with respect to the numerical simulation at the same high resolution.
Fig.~\ref{fig:relative_error} plots the resulting error variations over different time steps for the generalized synthesis case, as well as for different resolution ratios.
We also plot the error between coarse and fine simulations as a reference to better illustrate our synthesis accuracy.
Our synthesis error remains relatively small and bounded over time for these test cases.\\[-4mm]

\paragraph{Vortex structure preservation}
While our approach captures detailed vortex structures close to their fine simulation counterparts, our synthesis results sometimes exhibit crisper volutes, with less apparent diffusion in the rendered smoke compared to real physical simulations. Two reasons explain this behavior:
first, physical simulations can capture vortices with smaller scales which locally diffuse the smoke particles more than in the synthesized results; second, our network does not perfectly ensure spatial and temporal coherence between patches, and small mismatch between nearby patches can create local vorticity that attract smoke particles rather than diffuse them. One could add a local diffusion to simply counteract this effect; we kept all our results `as is', because a crisper look is, in fact, visually more attractive, and we also did not want to alter the results with post-processing in any way, which could obfuscate the interpretation of our results. It may also be noted that high-resolution structures are often synthesized even if the low-resolution simulation has seemingly not even any smoke in the area (see the trailing wisps in Fig.~\ref{fig:diff_res} or the rising plumes in Fig.~\ref{fig:re-simulation1}): as we synthesize a high-resolution velocity field directly rather than smoke density, low-resolution flows can have small velocity variations in regions where no smoke particles were driven towards, but our network has learned that these velocity configurations become, in fact, full-blown smoke structures at high resolution. \\[-4mm]

\paragraph{Synthesis over different resolutions}
We tried upsampling up to a ratio of 8 in each dimension, and obtained reasonable synthesis results as demonstrated in Fig.~\ref{fig:diff_res}. Note however that our learning framework does not currently support upsampling to arbitrary resolutions for a given training set: we can only upsample with the same resolution ratio as the training set we are given.\\[-4mm]

\subsection{Generalizability}

As described earlier, we classify our upsampling task into three types: a generalized synthesis, where after training from a varied set of simulation examples, smoke simulations can be quickly generated from a coarse input through ``interpolation'' (and moderate ``extrapolation'') from the training simulations; a restricted synthesis, where after training from a more restricted set of simulation examples, high-resolution smoke animations can be quickly generated from a coarse input through mostly ``interpolation'' of the training simulations; and finally, a re-simulation synthesis, where a given coarse-fine animation pair is used to train the network, and one can very quickly explore tiny changes in the coarse input to regenerate re-simulations. Obviously, the last type is most accurate when compared to the associated fine simulation, while the first one is likely to be the least accurate, particularly for high Reynolds numbers where turbulence is to be expected.
Several examples of upsampling we show in this paper were designed to illustrate how our approach can generalize a smoke flow based on the training set, and does not suffer from overfitting issues. 
The generalized synthesis of Fig.~\ref{fig:generalized_synthesis}, for instance, relies only on \emph{two} training simulations, one using a single sphere obstacle, and the second one using a set of 5 spheres placed on a common vertical plane. Synthesizing upsampled flows from these two sequences with significant variations of the initial conditions  (e.g., by either adding/removing ball obstacles, changing the obstacle shape, or rotating the 5-ball configuration by an angle) lead to visually plausible results. At the other end of the spectrum, even very restricted re-simulation examples in Figs.~\ref{fig:re-simulation1} and~\ref{fig:re-simulation2} exhibit some amount of both ``interpolation'' and ``extrapolation'' from the original simulation: recall that the training for re-simulation uses only \textit{a small subset} of all patches, sampled over time and space; so a synthesized re-simulation must rely on linear combinations of these patches, instead of directly replaying the fine animation.\\[-6mm]

\subsection{Comparison with other upsampling approaches}

There are very few previous works that can synthesize well high-resolution flow fields from low-resolution simulation inputs based on a neural network. 
The most relevant approach, proposed by Chu et al.~\shortcite{Chu-2017}, used a CNN-based feature descriptor to synthesize high-resolution smoke details, also based on a local patch-based synthesis scheme.
However, they rely on a nearest-neighbor search during synthesis, which vastly restricts the space of synthesized flow structures and makes the animation results often visually unnatural: smoke structures appear biased towards particular directions; for example, see Fig. 12 in their paper.

Another relevant recent work is tempoGAN network~\cite{Xie-2018}, which also targets high-resolution simulations from coarse inputs; however, unlike our work where the low-resolution input is obtained from coarse numerical simulations, they target the upsampling of a \emph{downsampled version of a high resolution flow simulation}, which has the advantage of keeping the main vortical structures mostly unchanged, just like in image upsampling. 
However, this is quite unreasonable for flow simulation upsampling: as we have demonstrated, turbulent flow structures often differ \emph{significantly} from their fine counterparts when computed on coarser resolutions, see Figs.~\ref{fig:high_low_example} and~\ref{fig:re-simulation1} for instance.
In addition, their method is much slower than ours: they require around 12 s. to synthesize a flow at a resolution of 200$\times$200$\times$200, while ours takes merely 3 s. (both were computed on GPU for fairness of evaluation). 

\begin{figure}[t]%\vspace*{-4mm}
	\centering
	\begin{subfigure}[b]{0.49\columnwidth}
		\includegraphics[width=\linewidth]{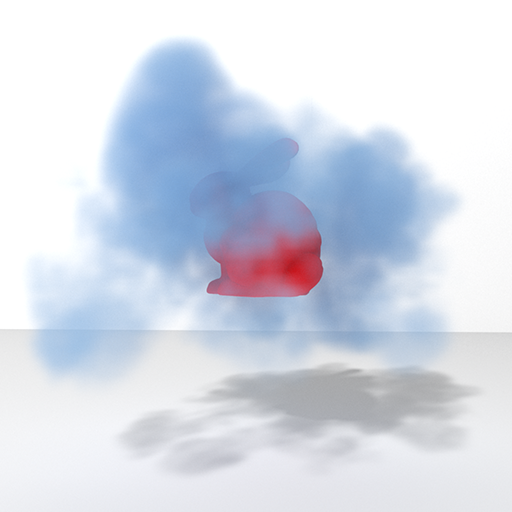}\vspace*{-2mm}
		\caption{Downsampled input}
		\label{fig:bunny-SR-downsampled}
	\end{subfigure}%
	\begin{subfigure}[b]{0.49\columnwidth}
		\includegraphics[width=\linewidth]{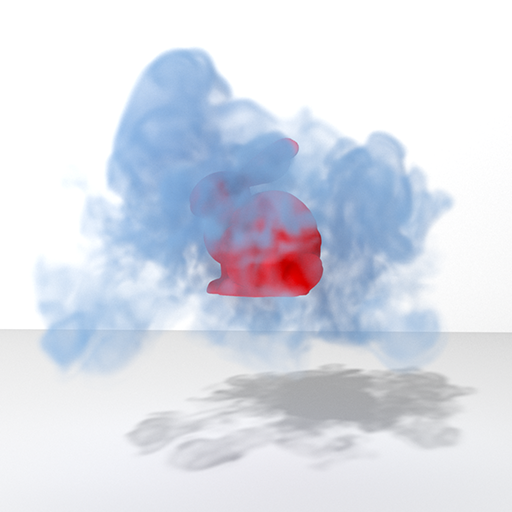}\vspace*{-2mm}
		\caption{tempoGAN result}
		\label{fig:bunny-SR-tempo}
	\end{subfigure}\\
	\begin{subfigure}[b]{0.49\columnwidth}
		\includegraphics[width=\linewidth]{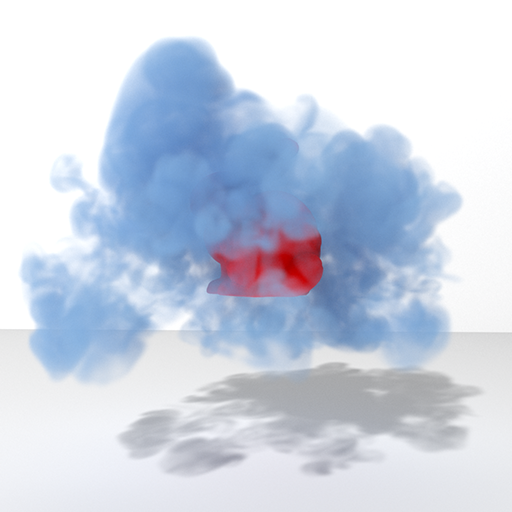}\vspace*{-2mm}
		\caption{Our result}
		\label{fig:bunny-SR-ours}
	\end{subfigure}%
	\begin{subfigure}[b]{0.49\columnwidth}
		\includegraphics[width=\linewidth]{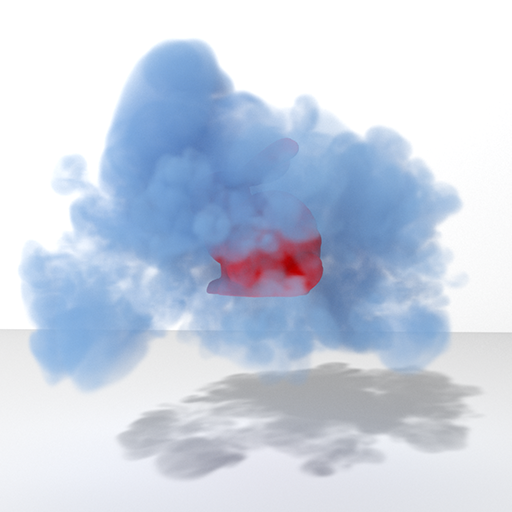}\vspace*{-2mm}
		\caption{High-resolution animation}
		\label{fig:bunny-SR-fine}
	\end{subfigure}%
	\vspace*{-3mm}
	\caption{ \textbf{From downsampled to upsampled flows.} From a smoke animation computed from a downsampled (4 times along each dimension) fine animation, tempoGAN~\protect\cite{Xie-2018} can make the smoke look sharper (b), but fail to capture the correct dynamics of the fine simulation (d); our approach, instead, captures the smoke animation much more closely.\vspace*{-2mm}}
	\label{fig:bunny_super-resolution}
\end{figure}

\begin{figure}[t]
	\centering
	\includegraphics[width=\columnwidth]{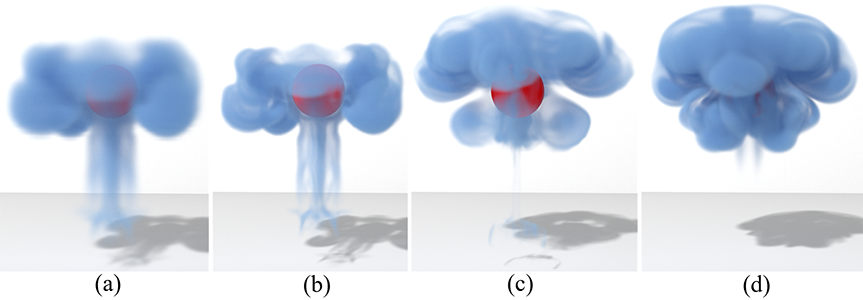}\vspace*{-4mm}
	\caption{\textbf{Comparison with tempoGAN network.} We perform upsampling comparison from a low-resolution numerical simulation input (50$\times$50$\times$50) between the tempoGAN network~\protect\cite{Xie-2018} and ours, both at the resolution of 200$\times$200$\times$200: (a) low-resolution numerical simulation input; (b) tempoGAN upsampling result; (c) our network upsampling result; (d) the ground-truth fine numerical simulation.\vspace*{-4mm}}
	\label{fig:comparison}
\end{figure}

Nevertheless, we tried to apply our approach using a downsampled version of a fine simulation in Fig.~\ref{fig:bunny_super-resolution} to offer a visual comparison between the tempoGAN network synthesis and ours. 
For a downsampled input (Fig.~\ref{fig:bunny_super-resolution}(a)), the tempoGAN network adds details to make the low-resolution smoke sharper (Fig.~\ref{fig:bunny_super-resolution}(b)), but as discussed in their paper, it may introduce undesirable structures; it may also be difficult to capture the vortical structures coming from the dynamic instabilities of the related high-resolution simulation (Fig.~\ref{fig:bunny_super-resolution}(d)).
In stark contrast, our network synthesis faithfully captures the main structure of the high-resolution flow (Fig.~\ref{fig:bunny_super-resolution}(c)), at a fraction of the execution time of tempoGAN.  
Now, if a \emph{coarse} numerical simulation is used as an input as in Fig.~\ref{fig:comparison}, our approach still significantly outperforms tempoGAN: even if our approach uses a single training simulation (of smoke rising over a simple sphere), a downsampled input of a flow around a bunny leads to an upsampled animation far closer to the real fine simulation than what tempoGAN offers.

\begin{figure}[t]%\vspace*{-4mm}
	\centering
	\includegraphics[width=\columnwidth]{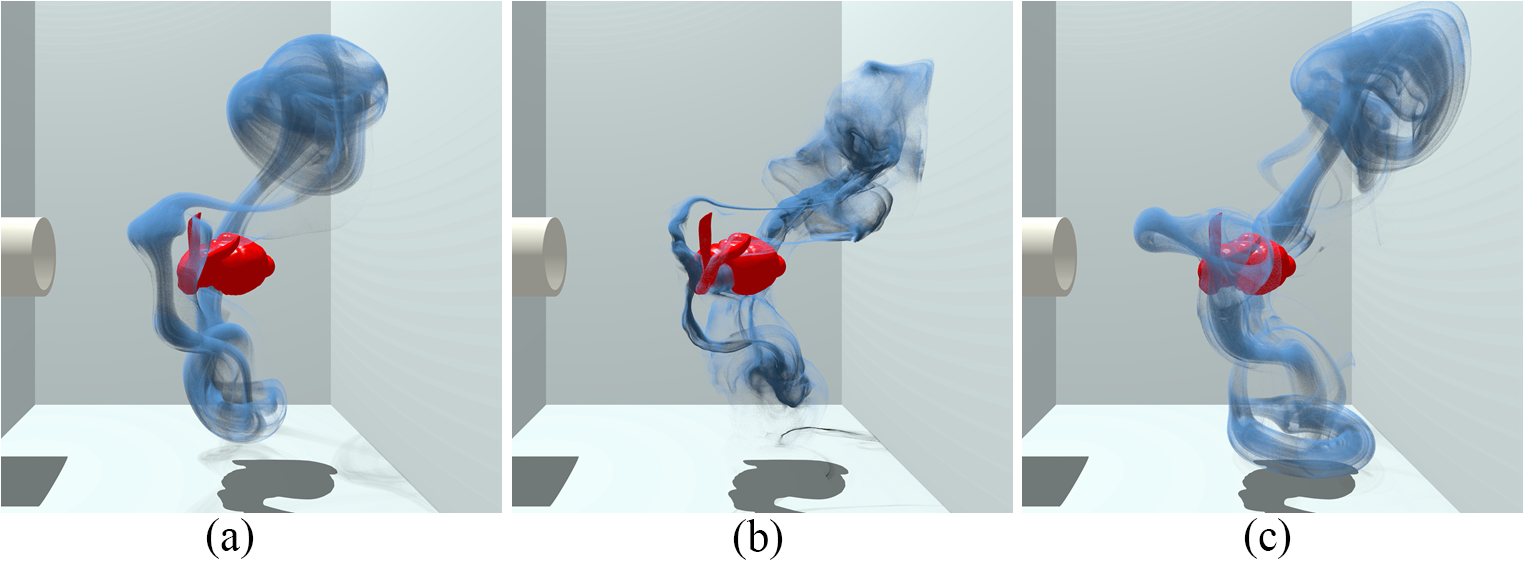}\vspace*{-3.5mm}
	\caption{\textbf{Potential limitations.} If we train our network on a flow over a sphere, upsampling a coarse animation of a turbulent flow over a bunny-shaped obstacle (a) can lead to significant inaccuracy (b) compared to the ground-truth solution (c): the training patches from the coarse simulation are simply not diverse enough to offer a good prediction. \vspace*{-2mm}}
	\label{fig:failure_case}
\end{figure}

\subsection{Limitations}
Our method is not without limitations, however. One cannot expect poorly-chosen training sets to provide predictive upsampling as we now detail to help understand what to expect from our approach. \\[-2mm]

First, our local patch approach cannot guarantee a perfect spatial and temporal coherence in the results --- although visual artifacts are all but impossible to notice in practice. 
Note that the patch coherence is strongly related to the generalization property of the network.
If a coarse input patch deviates significantly from the training patches, it will then be difficult to represent as a meaningful combination of training patches; the network behavior in this case is not quite predictable, and incoherence is likely to occur.
In such a case, nearby mismatched patches may create large velocity gradients (along with a strong vorticity) in the overlapped regions, which will attract smoke particles and result in very thin and unnatural smoke features --- see for instance Fig.~\ref{fig:failure_case}, where we train our network with a simple flow simulation around a ball obstacle, but synthesize a fast, turbulent flow around a bunny-shaped obstacle. More generally, very turbulent flows are simply difficult to upsample accurately:
since they are chaotic, an arbitrary coarse simulation may contain patches widely different from even a large sample of training patches.
Moreover, the difference between coarse and fine turbulent flows may increase exponentially over time, adding an additional difficulty for such fast flows. However, if the coarse inputs are downsampled versions of fine simulations like it was assumed in tempoGAN~\cite{Xie-2018}, these inputs are of course much more ``predictive'' of the motion even in the case of turbulent flows, and our approach indeed outperforms tempoGAN in this specific super-resolution case, see Fig.~\ref{fig:bunny_super-resolution}.\\[-4mm]

Second, since the high-frequency components are synthesized by our network without a strict enforcement of divergence-free condition, it also does not guarantee the incompressibility of our results; however, given that the coarse simulation is incompressible by construction and that we enforce a near divergence-free dictionary, the resulting high-resolution animation remains nearly incompressible.

Lastly, our method may require a large amount of training patches to produce accurate results, which causes longer training times, especially for the generalized synthesis.
How to train the network based on fewer samples is still worth further study.
In addition, our patch sampling used to generate training data does not guarantee that all important local behaviors are captured; as we discussed earlier, we believe that our importance sampling strategy could be further refined to improve generality.

\section{Conclusion}

In this paper, we proposed a dictionary-based approach to synthesizing high-resolution flows from low-resolution numerical simulations for the efficient (possibly iterative) design of smoke animation. In sharp contrast to previous works that only add high-frequencies through noise or fast procedural models, our approach learns to efficiently predict the appearance of fine details based on the results of coarse and fine offline numerical simulations.
A novel multiscale dictionary learning neural network is formulated based on a space-time encoding and a phase-space encoding of the flow, and then trained through a set of coarse and fine pairs of animation sequences. 
From any input coarse simulation, a high-resolution simulation can then be approximated via a sparse representation of the local patches of the input simulation by simply applying our trained network per patch, followed by a sparse linear combination of high-resolution residual patches blended into a high-resolution grid of velocity vectors. 
We also highlighted the key advantages of our method with respect to previous methods that either just added high-frequency noise or used a very limited space of upsampled patches, and provided a clear analysis of the possible failure cases for fast and turbulent flows. 
We now believe that our use of sparse combinations of patches from a well-chosen over-complete dictionary offers a rich basis for future neural-network based approaches to motion generation, not limited to smoke simulations.

\bibliographystyle{ACM-Reference-Format}
\bibliography{ss_dbnn}

\end{document}